\documentclass{sig-alternate-workshop}
\usepackage{amsmath}
\usepackage{url}
\usepackage{times}
\usepackage{epsfig}
\usepackage{subfigure}
\usepackage{xspace}
\usepackage{multirow}
\usepackage{color}
\graphicspath{{FIG/}}
\begin{document}

\newtheorem{theorem}{Theorem}
\newdef{definition}{Definition}
\newtheorem{proposition}[theorem]{Observation}
\newenvironment{proofsketch}[1][Proof Sketch]{\begin{trivlist}\item[\hskip \labelsep {\bfseries #1}]}{\end{trivlist}}
\newcommand{\xhdr}[1]{\vspace{1.7mm}\noindent{{\bf #1.}}}
\newcommand{\hide}[1]{}
\newcommand{\eg}{\emph{e.g.}}
\newcommand{\ie}{\emph{i.e.}}
\newcommand{\cag}{Community Affiliation Graph\xspace}
\newcommand{\agm}{AGM\xspace}
\newcommand{\agmtwo}{AGM2\xspace}
\newcommand{\agmlong}{Community-Affiliation Graph Model\xspace}
\newcommand{\areaks}{AreaKS\xspace}
\newcommand{\semihide}[1]{{\tiny #1}}
\newcommand{\rev}[1]{{\underline{** #1 **}}}
\newcommand{\reminder}[1]{\textcolor{red}{[#1]}}
\newcommand{\beq}{\begin{equation}}
\newcommand{\eeq}{\end{equation}}

\newcommand{\denselist}{ \itemsep -3pt\topsep-10pt\partopsep-10pt }

\conferenceinfo{The 6th SNA-KDD Workshop '12 (SNA-KDD'12),} {August 12, 2012, Beijing, China.}
\CopyrightYear{2012}
\crdata{978-1-4503-1544-9}
\clubpenalty=10000
\widowpenalty = 10000

\title{
Structure and Overlaps of Communities in Networks 
}

\numberofauthors{2}
\author{
\alignauthor Jaewon Yang\\
       \affaddr{Stanford University}\\
       \email{crucis@stanford.edu}
\alignauthor Jure Leskovec\\
       \affaddr{Stanford University}\\
       \email{jure@cs.stanford.edu}
}

\maketitle

\begin{abstract}

One of the main organizing principles in real-world social, information and technological networks is that of {\em network communities}, where sets of nodes organize into densely linked clusters.
Even though detection of such communities is of great interest, understanding the structure communities in large networks remains relatively limited. Due to unavailability of labeled ground-truth data it is practically impossible to evaluate and compare different models and notions of communities on a large scale.

In this paper we identify 6 large social, collaboration, and information networks where nodes {\em explicitly} state their community memberships. We define ground-truth communities by using these explicit memberships. We then empirically study how such ground-truth communities emerge in networks and how they overlap. We observe some surprising phenomena. First, ground-truth communities contain high-degree hub nodes that reside in community overlaps and link to most of the members of the community. Second, the overlaps of communities are more densely connected than the non-overlapping parts of communities, in contrast to the conventional wisdom that community overlaps are more sparsely connected than the communities themselves.

Existing models of network communities do not capture dense community overlaps.
We present the {\em \agmlong} (\agm), a conceptual model of network community structure, which reliably captures the overall structure of networks as well as the overlapping nature of network communities.

\noindent
{\bf Categories and Subject Descriptors:} H.2.8 {\bf [Database
Management]}: {Database Applications -- {\em Data mining}}

\noindent
{\bf General Terms:} Algorithms, theory, experimentation.

\vspace{-.2cm}

\end{abstract}

\section{Introduction}
\label{sec:intro}

Nodes in networks organize into densely linked groups that are commonly referred to as {\em network communities}, clusters or modules~\cite{fortunato09community,papadopoulos11community}. Studying networks at the level of communities is very useful as there are many reasons why social, information and technological networks organize into communities. For example, society is organized into groups, families, friendship circles, villages and associations~\cite{feld86focused,simmel64affiliations}. On the World Wide Web, topically related pages may link more densely among themselves~\cite{flake00_efficient}. 


Even though extracting network communities is a fundamental problem, understanding of the networks at the level of network communities has been relatively limited due to several challenges~\cite{RCCLP04_PNAS,jure08ncp2}. 
There exists a large number of different definitions and models of network communities~\cite{fortunato09community,RCCLP04_PNAS}, and many formalizations of community detection lead to intractable NP-hard optimization problems~\cite{Schaeffer07_survey}.  
Moreover, the lack of reliable ground-truth makes the evaluation of such models extremely difficult.

\xhdr{Present work}
In this work we study the connectivity structure of ground-truth communities in order to develop models for network communities. We identified a set of 6 different large social, collaboration, and information networks where we can reliably define the notion of {\em ground-truth communities}. Networks we study come from a number of different domains and research areas. In all these networks nodes explicitly state their ground-truth community memberships. The size of the networks ranges from hundreds of thousand to hundreds of millions of nodes and edges. The networks represent a wide range of edge densities, numbers of ground-truth communities, as well as sizes and amounts of community overlap. The availability of reliable ground-truth communities has a profound effect. It allows us to quantify the structure of ground-truth communities and build better models of how nodes organize into communities.

We study how ground-truth communities of nodes connect inside the community, how they connect to the rest of the network, and how they overlap. This way we can empirically study on a large scale how real communities map on the underlying social network structure, and how real communities overlap and interact.

We show the following for a broad range of networks across diverse domains: First, communities in social networks contain high-degree connector nodes that have an edge to most of the community members. Second, the overlaps between communities are more densely connected than the non-overlapping parts. We view the second of these findings as particularly surprising as it goes against the conventional wisdom that communities (both overlapping and non-overlapping) are more densely connected than their boundaries or the overlaps themselves~\cite{Ahn10LinkCommunitiesNature,palla05_OveralpNature}. Thus, rather than shedding light on the debate on whether and how much communities overlap in networks, our findings suggest the need to revisit standard models of community structure to account for the fact that nodes in the overlap of communities are more likely to be connected.

What underlying process causes community overlaps to be denser than the communities themselves? This question motivates the second main contribution of this work: We present a family of probabilistic generative models for graphs that capture the observed phenomena and produce graphs with realistic community structure. We build on models of affiliation networks~\cite{breiger74groups,Lattanzi09AffiliationNetworks} and develop the {\em \agmlong} (\agm) which reliably reproduces the organization of networks into communities and the overlapping community structure. In the affiliation network, memberships of nodes to communities are modeled with a bipartite graph, where on the ``left'' we have the nodes of the social network, and on the ``right'' are the nodes representing communities. The edges of this bipartite graph model node-community affiliations. The central idea in generating social networks  based on the affiliation network is that links among people stem from one or more common or shared community affiliations~\cite{breiger74groups}.

In our model communities arise due to shared group affiliations~\cite{breiger74groups,simmel64affiliations,feld86focused}. We model the probability of an edge between a pair of nodes as a function of the communities that the two nodes share. Community assignments in our model are probabilistic which allows for flexibility in the structure of community overlaps. We mathematically analyze  the \agm and obtain rigorous results showing that the model leads to community structure and overlaps as observed in real data. Experiments on a range of network datasets establish that the \agm reliably captures node community memberships, internal structure of the groups and generates realistic group overlaps.


Overall, our work has three main contributions:
\begin{itemize}
\denselist
    \item Identification of networks with explicit notion of  ground-truth communities.
    \item Novel observation that community overlaps are densely connected.
    \item \agmlong that explains the emergence of dense community overlaps and accurately models network community structure.
\end{itemize}
Our results have implications in several contexts:
\begin{itemize}
\denselist
  \item {\em Design of new community detection methods:}  Nearly all community detection methods assume sparse community overlaps~\cite{palla05_OveralpNature,Ahn10LinkCommunitiesNature}. This means that these methods cannot properly detect communities in large networks -- they would either mistakenly identify the overlap as a separate community or merge two overlapping communities into a single community. Thus, our findings have important implications for the development of new network community detection methods.
  \item {\em Evaluation based on ground-truth communities:} Our identification of networks with explicit ground-truth communities allows for quantitative evaluation: based on ground-truth communities we can evaluate the accuracy, \ie, what fraction of the members of the ground-truth community a particular method identified.
  \item {\em Synthetic benchmarks:} Our model can be used to generate synthetic benchmark datasets for evaluation and analysis of network community detection methods.
\end{itemize}

\section{Related Work}
\label{sec:related}


It is important to note the fundamental contrast between one of our main findings here --- that the community overlaps are denser than communities themselves --- and a massive body of work on network community detection (\ie, unsupervised graph clustering problem of inferring communities; see~\cite{luxburg05_survey,Schaeffer07_survey,fortunato09community} for surveys of this area). While community detection work seeks to infer potential communities in a network based on density of linkage, we start from the other end of the problem. We start with a network in which the communities have already been explicitly identified and seek to model their structure. Thus, our work here is directly relevant for community detection as it identifies the properties of real communities and a presents a realistic model. A natural next step (that we do not address here) is then to fit the model to a given graph and identify communities.

We also note that the finding that community overlaps are denser than communities themselves nicely extends the notion of homophily in networks~\cite{mcpherson01homophily}. The `strength of weak ties'~\cite{granovetter73ties} and small-world models~\cite{watts98collective} lead to the idea that homophily in networks operates in small pockets where inside the pocket nodes link strongly among themselves, and weakly to other pockets. In this respect our work here represents an extension to the understanding of homophily. In a sense we are discovering {\em pluralistic homophily}\footnote{We thank Michael Macy for coining this term for us.} where the similarity of one node to another is the number of shared affiliations, not just their similarity along a single dimension. This view of tie formation is consistent with the works of Simmel~\cite{simmel64affiliations} on the web of affiliations, and Feld~\cite{feld86focused} on focused organization of social ties. In both views networks compose of overlapping tiles or social circles that serve as organizing principles of nodes in networks.



There has also been considerable work on probabilistic models for graph generation. The discovery of degree power-laws and other properties of static and dynamic graphs led to the development of random graph models that exhibited such properties~\cite{kumar00stochastic,Lattanzi09AffiliationNetworks,jure07evolution,watts98collective}.
See~\cite{mitzenmacher04brief,chakrabarti06survey} for surveys of this area. The main difference here is that our goals are more ambitious as we aim to accurately model both the overall network structure as well as the node community memberships and the overlaps of communities.

Our \agmlong (\agm), which produces realistic graphs as well as community overlaps, is an example of a bipartite affiliation network model~\cite{breiger74groups,Lattanzi09AffiliationNetworks,zheleva09affiliation}. Affiliation networks have been extensively studied in sociology~\cite{breiger74groups} as a metaphor of classical social theory concerning the intersection of persons with groups, where it has been recognized that communities arise due to shared group affiliations~\cite{breiger74groups,simmel64affiliations}. In affiliation network models, nodes of the social network are affiliated with communities that they belong to and the links of the underlying social network are derived based on the community affiliation network. The most related to our model is the work of Lattanzi and Sivakumar~\cite{Lattanzi09AffiliationNetworks} who studied the macroscopic evolution of networks and proposed an affiliation network model for social networks. They proved that networks arising from the model exhibit power-law degree distributions, densification power law and shrinking diameter. There is a small but crucial difference between the two models. \cite{Lattanzi09AffiliationNetworks} posed a model where edge creation probability {\em decreases} with community size. We relax this assumption and allow \agm communities to have arbitrary edge probabilities. This way \agmlong acquires the necessary flexibility to accurately model the community structure of real-world networks.

\section{Dataset description}
\label{sec:data}
We aim to identify networks where nodes explicitly state their community memberships. Ideally such ground-truth communities would be real communities with shared values, mutual influence and common purpose or function. We consider a set of 6 large social, collaboration and information networks, where for each network we identify a graph and a set of ground-truth communities. We identify networks where nodes explicitly state their ground-truth community memberships. Members of these ground-truth communities share properties or attributes, common purpose or function. We did our best to identify networks in which such ground-truth communities can be reliably defined and identified.  Networks that we study come from a variety of domains. Their size ranges from hundreds of thousand to hundreds of millions of nodes and billions of edges.

\begin{figure}[t]
\centering
  \includegraphics[width=0.23\textwidth]{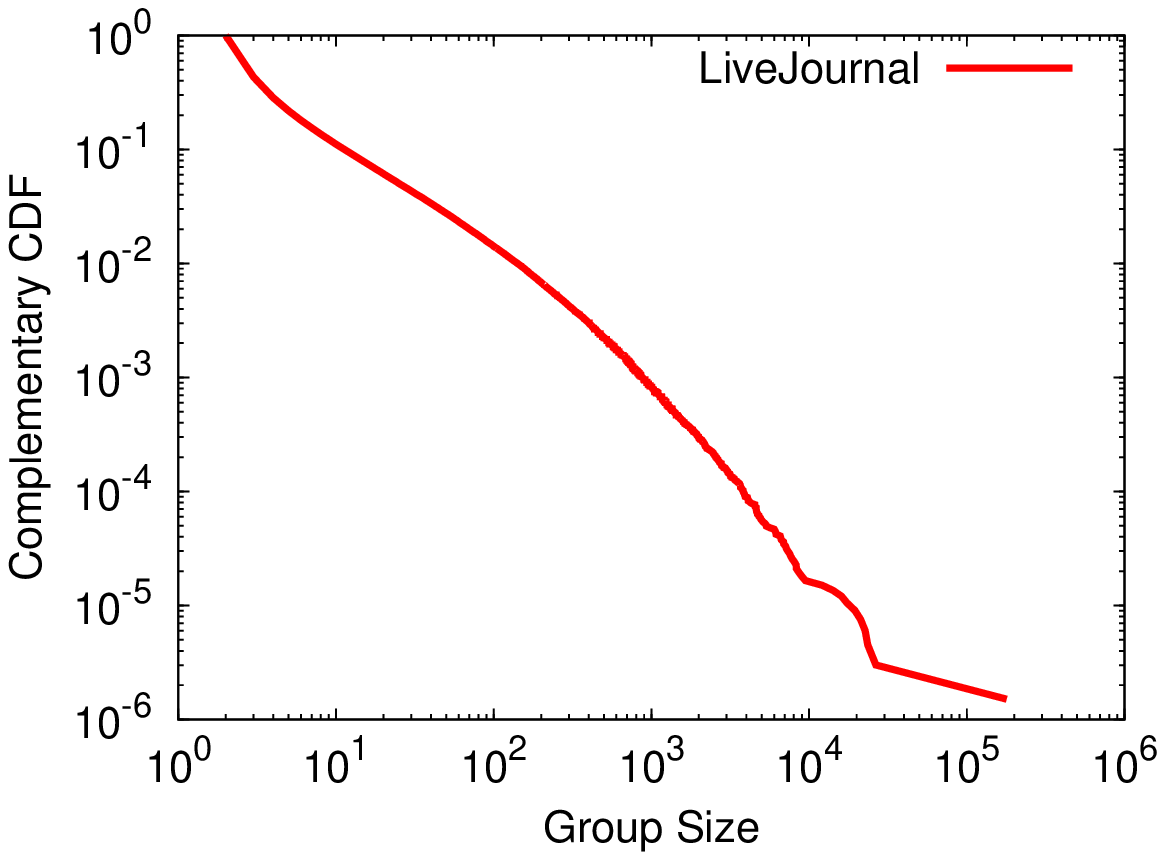}
  \includegraphics[width=0.23\textwidth]{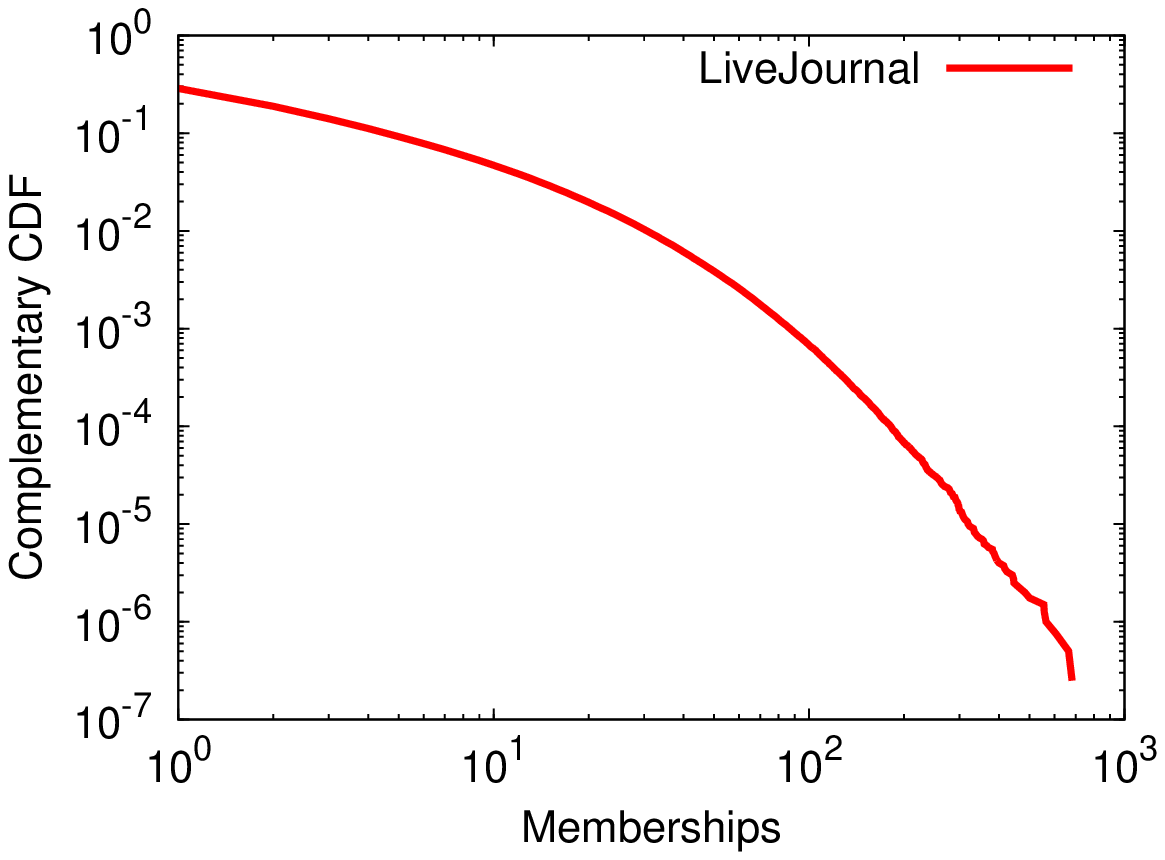}
  \caption{LiveJournal group statistics: (a) Group size distribution, (b) Number of group memberships per node. Other networks have similar behavior (not shown for brevity) \cite{jaewon11community}.}
  \label{fig:ljGroups}
\end{figure}

First we consider online social networks (the LiveJournal blogging community \cite{lars06groups}, the Friendster online network \cite{mislove07measurement}, and the Orkut social network \cite{mislove07measurement}) where users create explicit groups which other users then join.  Such groups serve as organizing principles of nodes in social networks and are focused on specific topics, interests, hobbies, affiliations, and geographical regions. Communities range from small to very large and are created based on specific topics, interests, hobbies and geographical regions. For example, LiveJournal categorizes communities into the following types: culture, entertainment, expression, fandom, life/style, life/support, gaming, sports, student life and technology. For example, there are over 100 communities with `Stanford' in their name, and they range from communities based around different classes, student ethnic communities, departments, activity and interest based groups, varsity teams, etc. Overall, there are over three hundred thousand explicitly defined communities in LiveJournal. Figure~\ref{fig:ljGroups} gives the statistics of group sizes and the number group memberships of nodes in LiveJournal. Similarly, users in Friendster as well as in Orkut define topic-based communities that others then join. Both networks have more than a million explicitly defined groups and each user can join to one or more such groups. We consider each group as a ground-truth community.

The second type of network data we consider is the Amazon product co-purchasing network~\cite{jure07viral}. The nodes of the network represent products and edges link commonly co-purchased products. Each product (\ie, node) belongs to one or more hierarchically organized product categories and products from the same category define a group which we view as a ground-truth community. This means members of the same community share a common function or role, and each level of the product hierarchy defines a set of hierarchically nested and overlapping communities.

Finally, we also consider the collaboration networks of DBLP and IMDB~\cite{lars06groups} where nodes represent authors/actors and edges connect nodes that have co-authored a paper/co-appeared in the movie. In DBLP we use publication venues as ground-truth communities which serve as proxies for highly overlapping scientific communities around which the network then organizes. In IMDB we found that language is a good way of defining ground-truth communities.

\begin{table}[t]
\centering
\tiny
    \begin{tabular}{l||r|r|r|r|r}
    Dataset & $N$ & $E$ & $C$ & $S$ & $A$ \\ \hline \hline
    LiveJournal&4.0 M&34.9 M&311,782&40.06&3.09\\
    Friendster&117 M&2,586.1 M&1,449,666&26.72&0.33\\
    Orkut&3.0 M&117.2 M&8,455,253&34.86&95.93\\
    DBLP&0.4 M&1.3 M&2,547&429.79&2.57\\
    IMDB&1.3 M&39.8 M&205&6,688.78&1.00\\
    Amazon&0.3 M&0.9 M&49,732&99.86&14.83\\
    \end{tabular}
    \caption{Dataset statistics. $N$: number of nodes, $E$: number of edges, $C$: number of communities, $S$: average community size, $A$: community memberships per node. M denotes one million.}
    \label{tab:data}
  \vspace{-3mm}
\end{table}

Table~\ref{tab:data} gives the dataset statistics. Observe that the size of networks ranges between hundreds of thousands to hundreds of millions of nodes and billions of edges. The number of ground-truth communities varies from hundreds to millions and there is also a nice range in group sizes and the node membership distribution.

All our networks are complete and publicly available: LiveJournal \cite{lars06groups}, Friendster \cite{mislove07measurement}, Orkut \cite{mislove07measurement}, Amazon \cite{jure07viral}, DBLP \cite{lars06groups} and IMDB \cite{lars06groups}.\footnote{All networks and the corresponding ground-truth communities are available at \url{http://snap.stanford.edu/data}} For each of these networks we identified a sensible way of defining ground-truth communities that serve as organizational units of these networks.
Note that we are careful to define ground-truth communities based on common affiliation, social circle, role, activity, interest, function, or some other property around which networks organize into communities~\cite{feld86focused,granovetter73ties}. 

Even though our networks come from very different domains and have very different motivation for formation of communities the results we present here are consistent and robust. Our work is consistent with the premise that is implicit in all network community literature: members of real communities share some (latent/unobserved) property or affiliation that serves as an organizing principle of the nodes and makes them well-connected in the network. Here we use these groups around which communities organize to explicitly define ground-truth. And, as we will later see, the ground-truth communities exhibit connectivity patterns that match our intuition of communities as densely connected sets of nodes.

\xhdr{Data preprocessing}
To represent all networks in a consistent way we drop edge directions and consider each network as an unweighted undirected static graph. Because members of the group may be disconnected in the network, we consider each connected component of the group as a separate ground-truth community. However, we allow ground-truth communities to be nested and to overlap (\ie, a node can be a member of multiple groups at once).

\section{Structure of communities}
\label{sec:experiments}

We now proceed to discuss our empirical findings that motivate the model we later develop.
We focus our analyses on two aspects of connectivity structure of ground-truth communities:
First, we investigate the connectivity properties of communities, \ie, we study the structure of induced graphs on the set of community members $S$. Second, we study  connectivity patterns of community overlaps and investigate the amount of edge clustering in the overlap versus the clustering in the non-overlapping parts of the community.



\begin{figure}[t]
\centering
  \subfigure[Edges inside the community]{\label{fig:sz.vol}\includegraphics[width=0.23\textwidth]{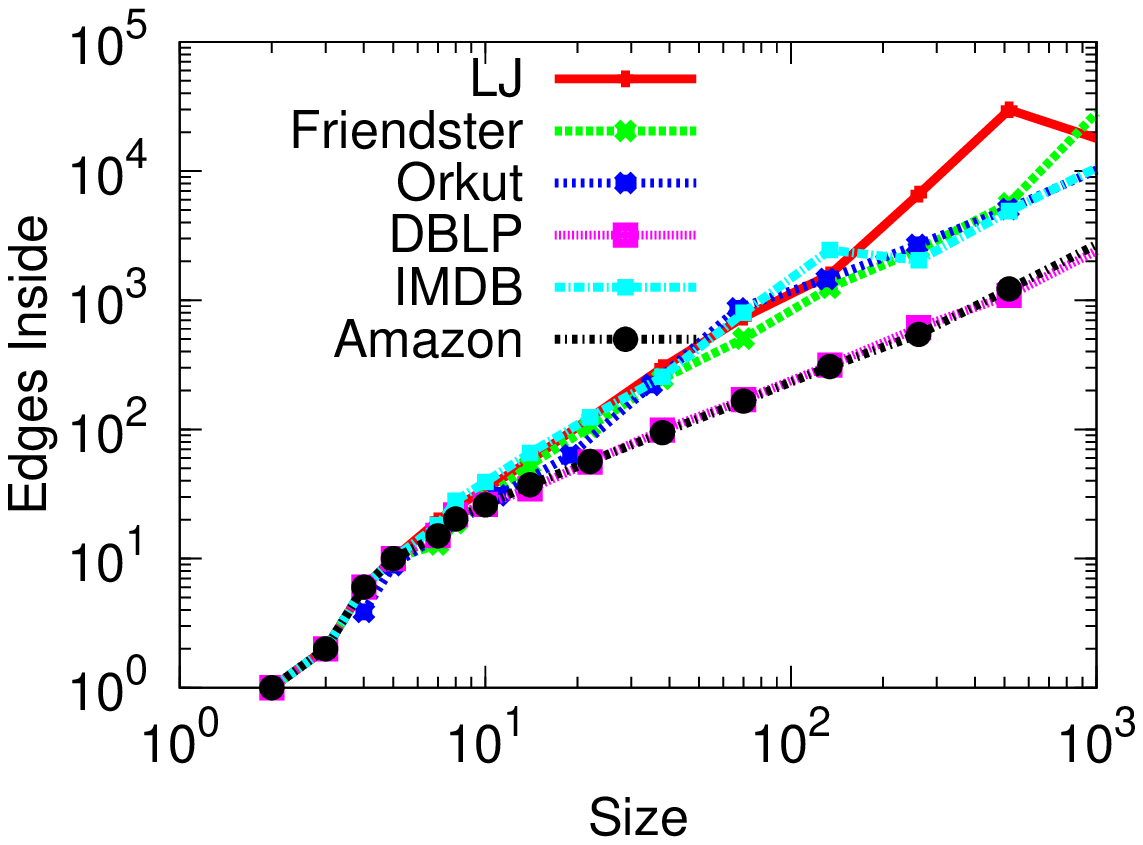}}
  \subfigure[Maximal ICDF]{\label{fig:sz.maxintdeg}\includegraphics[width=0.23\textwidth]{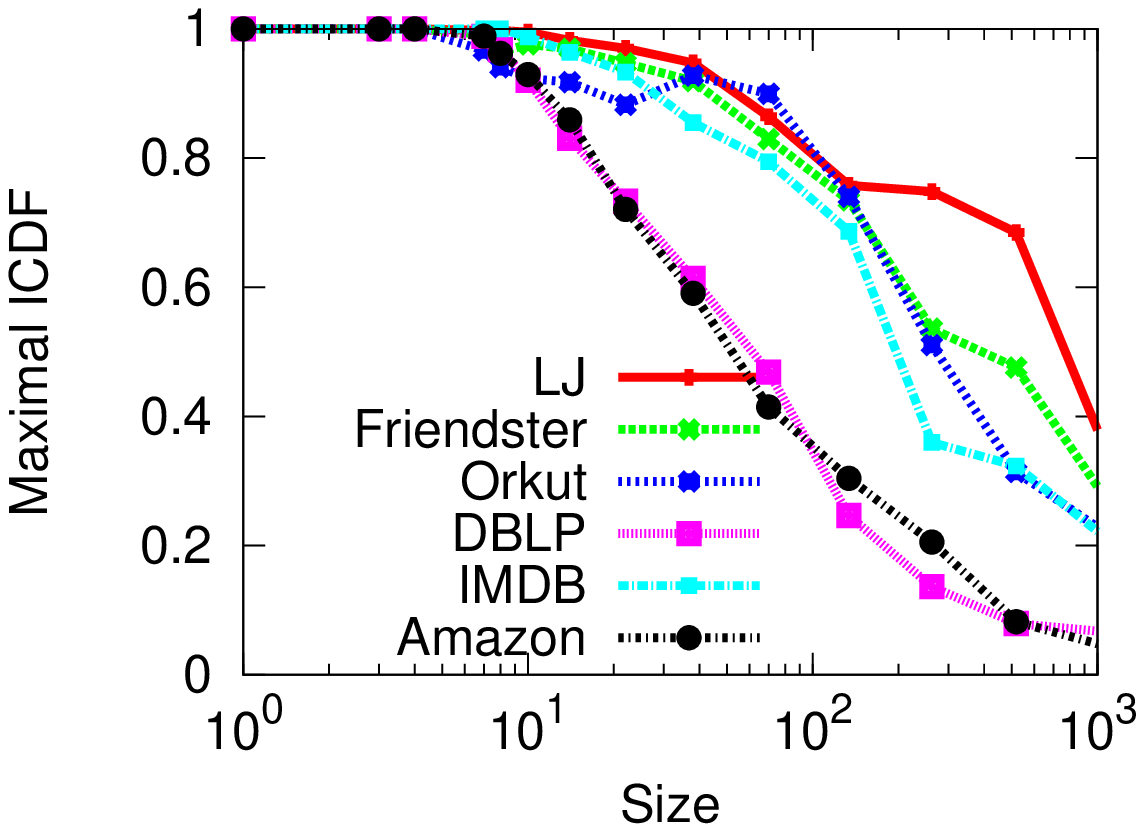}}
  \vspace{-2mm}
  \caption{(a) Edges inside the community, and (b) Maximal Internal Community Degree Fraction as a function of the community size.}
  \vspace{-3mm}
  \label{fig:group.stat}
\end{figure}

\begin{figure}[t]
\centering
  \subfigure[Overlap]{\label{fig:overlap.toy}\includegraphics[width=0.20\textwidth]{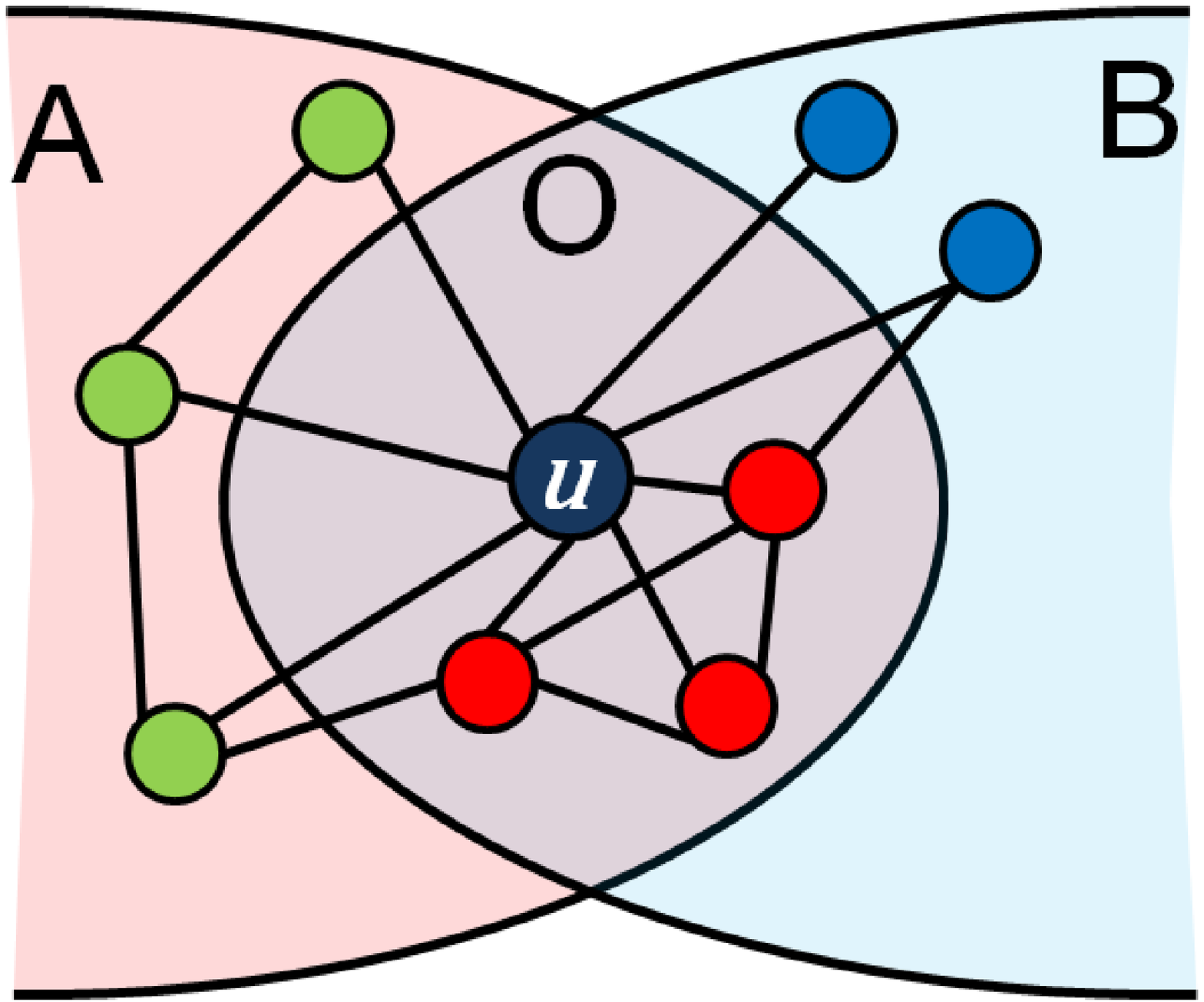}}
  \subfigure[Community Affiliation Network]{\label{fig:AGM.bipartite}\includegraphics[width=0.25\textwidth]{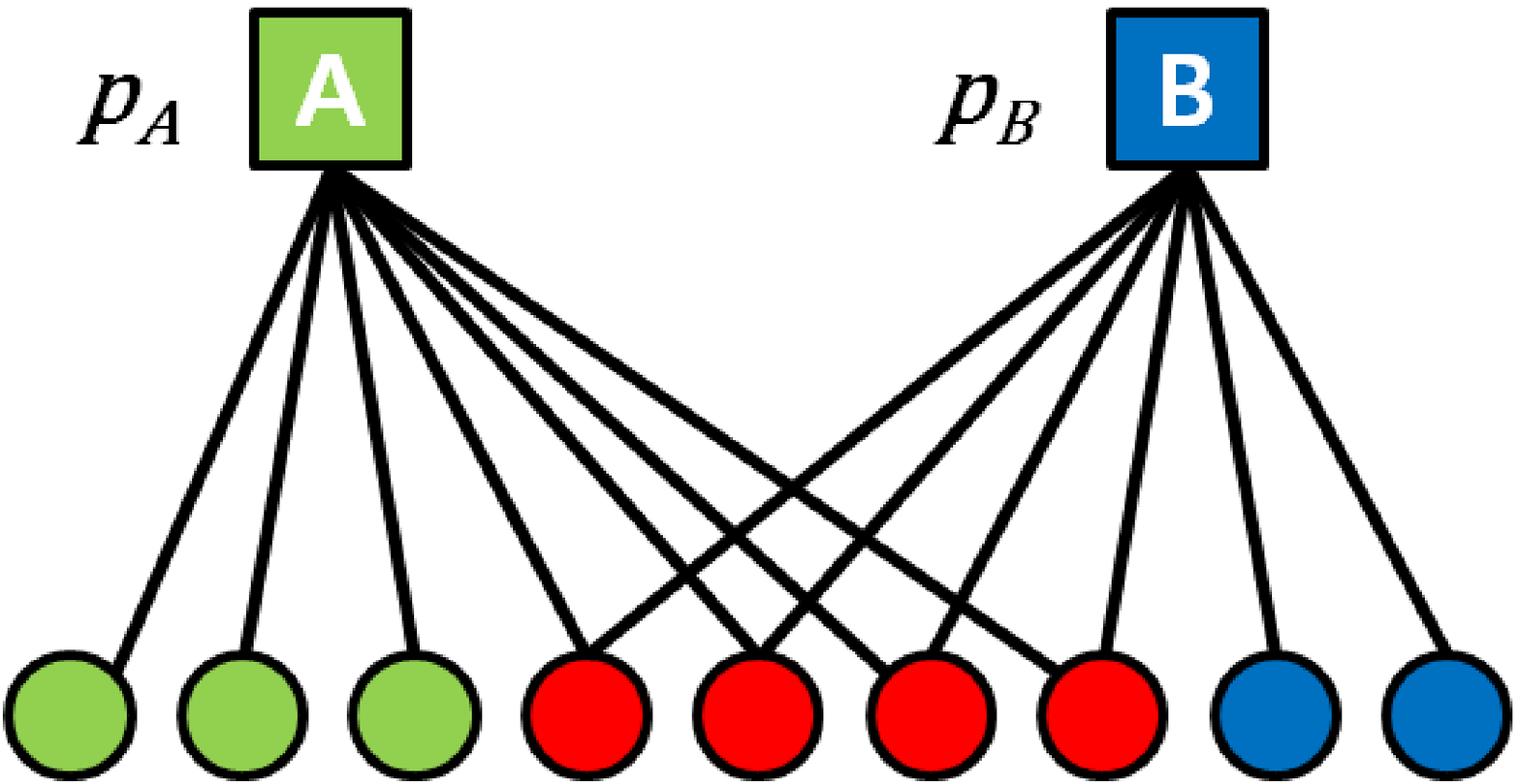}}
  \vspace{-2mm}
  \caption{(a) Overlap O of communities A and B.
  (b) Community Affiliation Network constructed from (a). }
  \label{fig:overlap.illustration}
  \vspace{-2mm}
\end{figure}


\xhdr{Connectivity of communities}
We first examine the relation between the size of the community (\ie, $|S|$) and the number of edges between the nodes of the community ($E_S=\{(u,v) | u,v \in S, (u,v)\in E)\})$. Figure~\ref{fig:sz.vol} shows the relation. Interestingly, across the range of datasets we consistently observe a form of the {\em Densification Power Law}~\cite{jure05dpl} where the number of edges in the community increases superlinearly with the community size, $|E_S| \propto |S|^\alpha$. We observe that all three online social networks (Orkut, LiveJournal and Friendster) have densification exponent $\alpha \approx 1.5$. We also note a similar exponent for IMDB. On the other hand DBLP and Amazon have lower value of $\alpha \approx 1.1$. However, it is important to correctly interpret these findings. Even though the absolute number of one's friends that are in the community increases with the size of the community (\ie, the number of edges increases superlinearly with the number of nodes), the {\em fraction} of community members whom one is friends with (\ie, density of the community) {\em decreases} with the community size (since $\alpha<2$). This suggests that when the community is small, its members build relationships among themselves, whereas in a large community, members are less embedded into the community.

The distinction between small and large communities becomes even clearer once we examine the existence of a connector/hub node in the community. To investigate this, we first define the {\em Internal Degree (ID)} $d_{in}(u,S)$ of node $u$ in a community $S$ to be the number of members of $S$ to which $u$ is connected, $d_{in}(u,S)=|\{v | v \in S, (u,v)\in E_S\}|$.
Then we define the {\em Maximal Internal Community Degree Fraction (ICDF)} $f_{in}(S)$ of community $S$ to be the maximal fraction of community members any member node is connected to, $f_{in}(S) = \max_{u\in S} d_{in}(u,S)/|S|$. For example, a Maximal ICDF of 0.7 of community $S$ means that there exists a node $u\in S$ that is connected to 70\% of all of the members of $S$.

Figure~\ref{fig:sz.maxintdeg} plots the average Maximal ICDF as a function of community size. We observe that in communities smaller than $\approx$100 nodes, there exist connector nodes that link to more than 80\% of all the community members. However, as the community size increases beyond 100, the Maximal ICDF tends to quickly decrease. This is interesting as it suggests that smaller communities tend to organize themselves around connector node(s) and thus share common bonds. There are two exceptions to this both of which can be nicely explained. Amazon is a product co-purchasing network and thus there is no internal reason why such connector nodes should exist. And the presence of a connector node in the DBLP network would suggest existence of special publication venues where the connector node would co-author all the papers, which again is not realistic.

\begin{figure*}[!t]
	\centering
    \subfigure[LiveJournal]{\label{fig:edge_prob.lj}\includegraphics[width=0.31\textwidth]{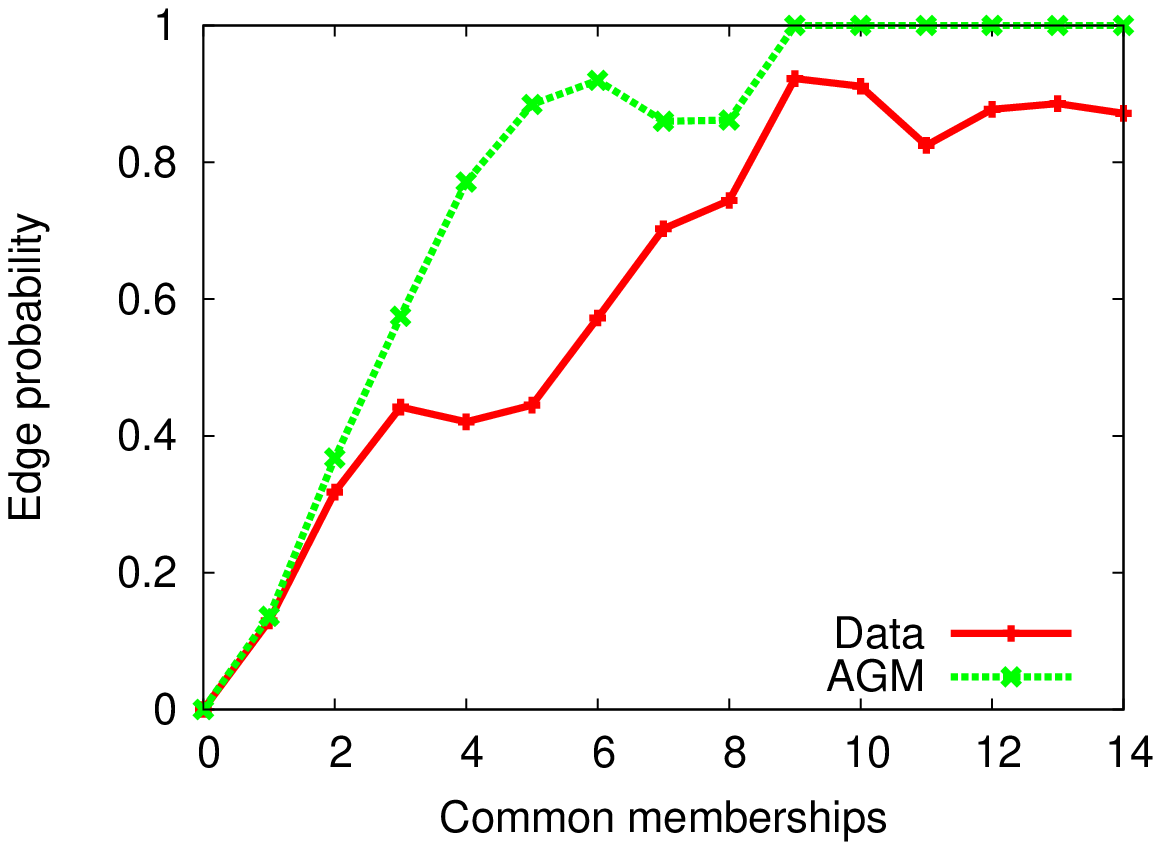}}
	\subfigure[Friendster]{\includegraphics[width=0.31\textwidth]{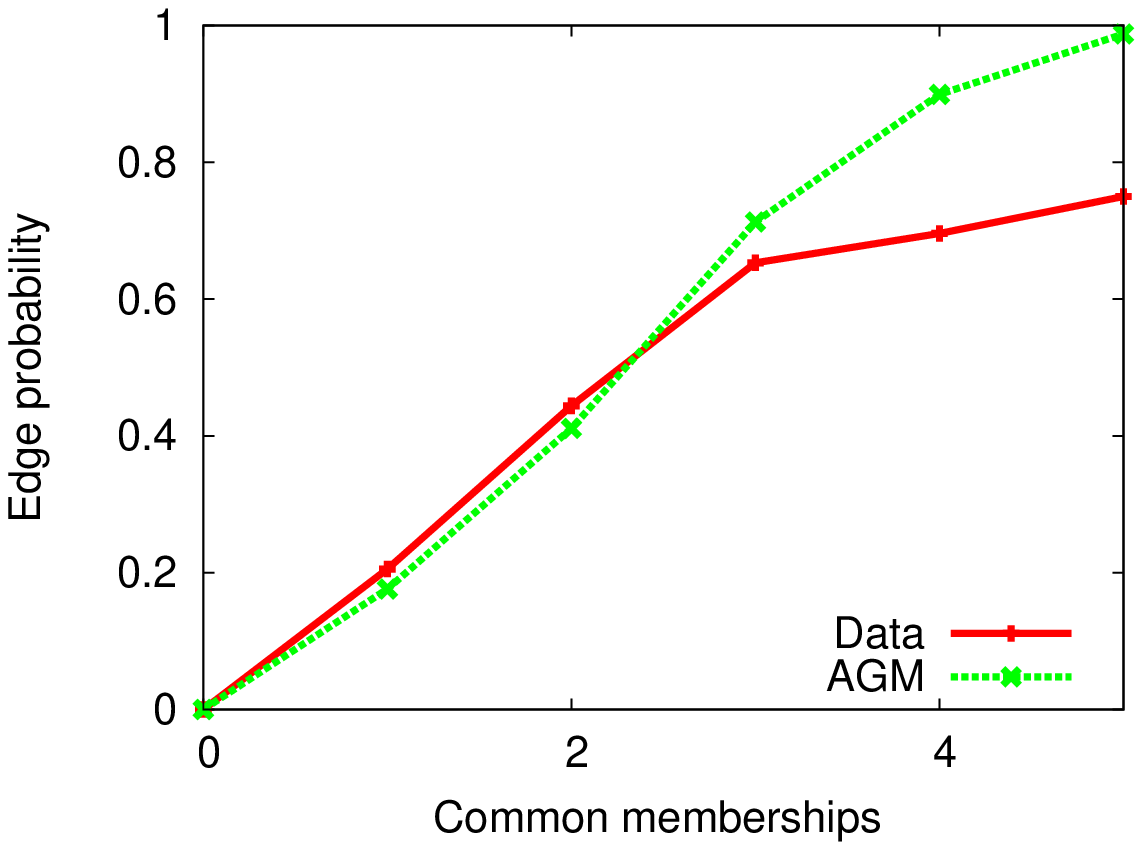}}
	\subfigure[Orkut]{\includegraphics[width=0.31\textwidth]{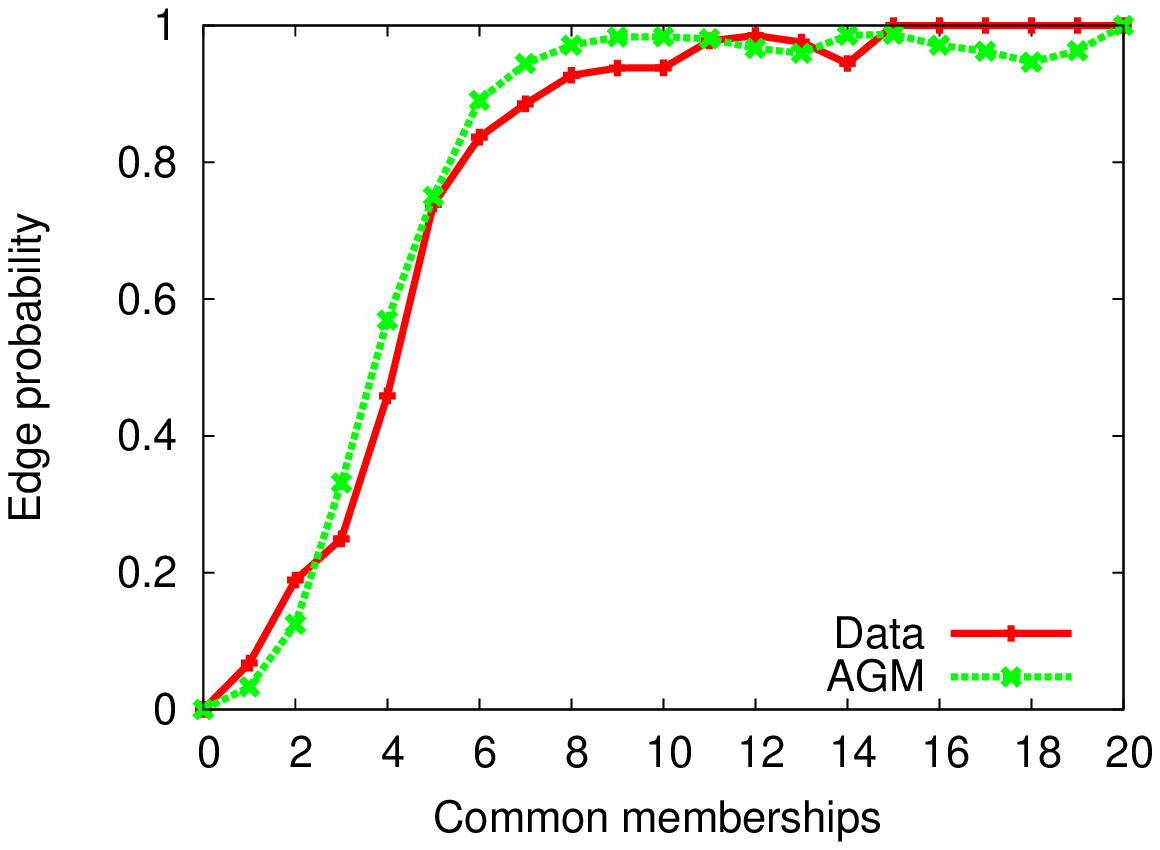}}
	 \subfigure[DBLP]{\label{fig:edge_prob.dblp}\includegraphics[width=0.31\textwidth]{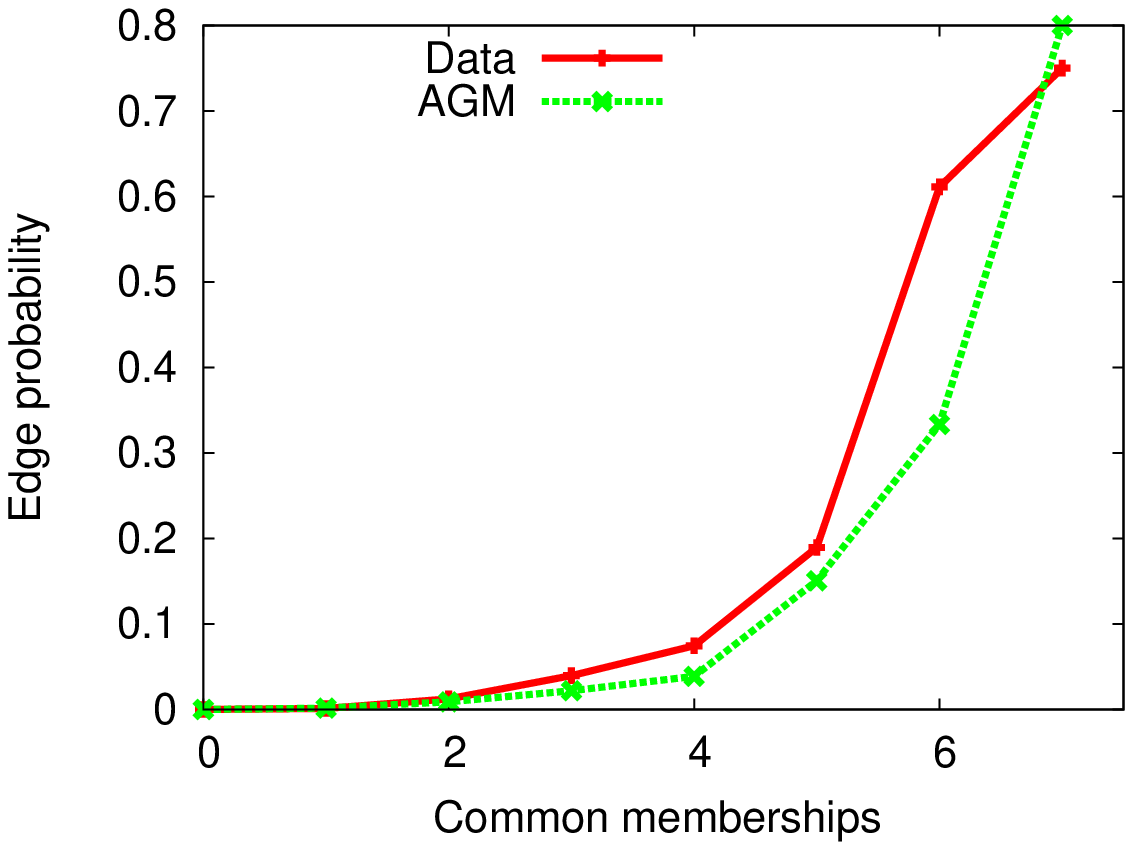}}
	\subfigure[IMDB]{\includegraphics[width=0.31\textwidth]{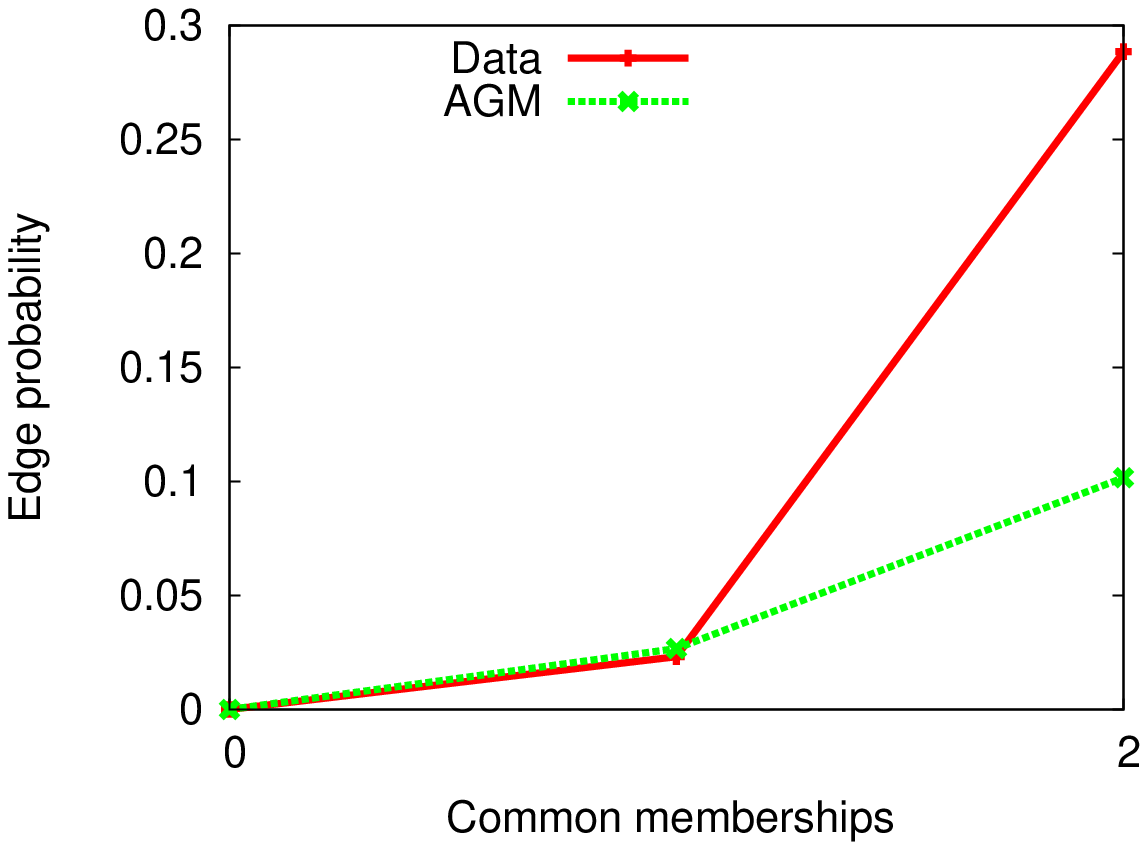}}
    \subfigure[Amazon]{\label{fig:edge_prob.amazon}\includegraphics[width=0.31\textwidth]{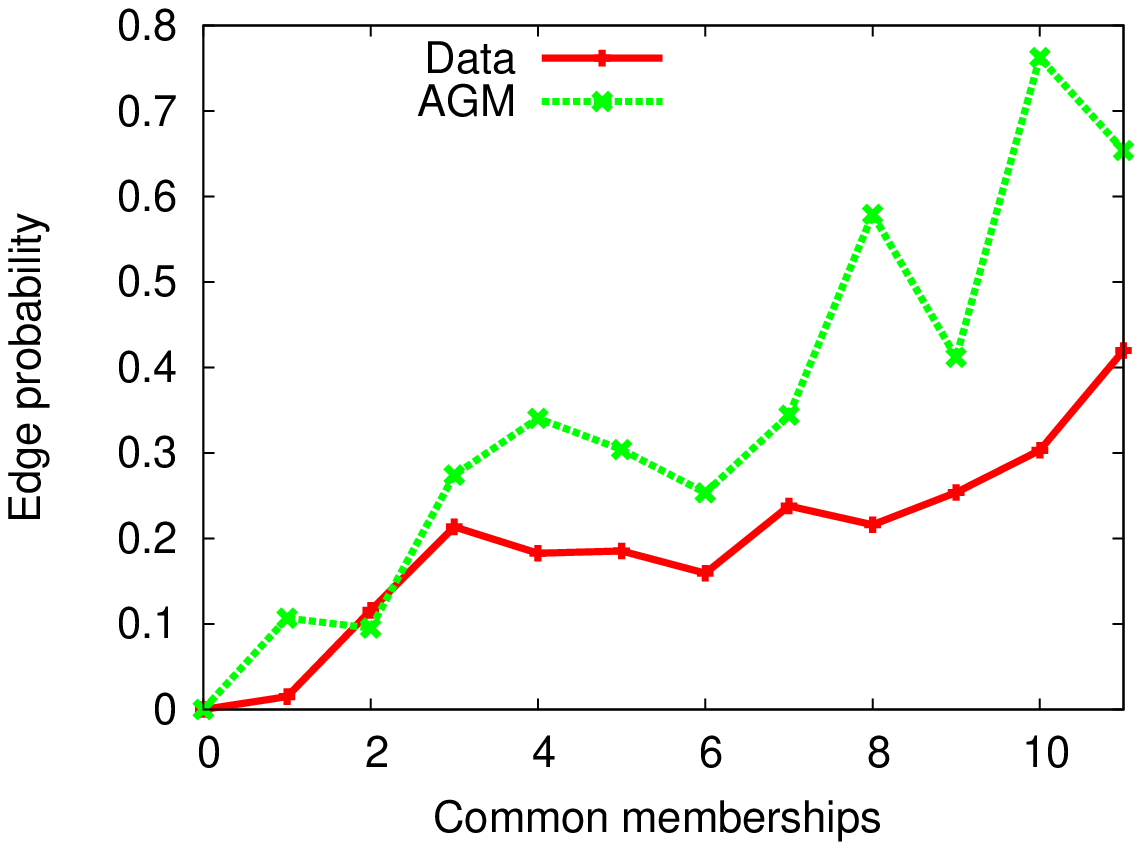}}
\vspace{-2mm}
	\caption{Edge probability between two nodes given the number of common communities that they belong to.}
\vspace{-0mm}
\label{fig:edge_prob}
\end{figure*}

\xhdr{Edge probability as a function of shared communities}
Communities in networks form overlaps when there exist nodes that belong to multiple communities. Figure~\ref{fig:overlap.toy} illustrates the setting of two communities $A$ and $B$, and nodes that belong to both communities reside in the {\em overlap} $O$. We study the structure of group overlap by simply asking what is the probability that a pair of nodes is connected if they share $k$ common community memberships, \ie, the nodes belong to the overlap of $k$ communities. Figure~\ref{fig:edge_prob} plots this probability (the red `Data' curve) for all six datasets. The figure also plots (the green `AGM' curve) the same quantity as modeled by our \agmlong that we will describe in the next section.

First, notice that all curves are generally increasing. This means that, the more communities a pair of nodes has in common, the higher the probability of an edge. In LiveJournal, for example, if a pair of nodes has 8 groups in common, the probability of friendship is nearly 80\%. To appreciate how strong the effect of shared communities is on the edge probability, one has to note that all our networks are extremely sparse. The background probability of a random pair of nodes being connected is $\approx 10^{-5}$, while as soon as a pair of nodes shares two communities, their probability of linking increases by 4 orders of magnitude (from $10^{-5}$ to $10^{-1}$).  We note that all other data sets have similar shapes --- the probability of a pair of nodes being connected approaches 1 as the number of common communities increases. While in online social networks the edge probability exhibits a diminishing-returns-like growth, in other datasets (IMDB, DBLP, Amazon) it appears to follow a threshold-like behavior.


In retrospective the above result is very intuitive. For example, in the context of social networks two students that belong to both a Tuesday salsa club and a Sunday Movie club are more likely to meet each other than if they would belong to only a single club. Thus, the more communities nodes share, the more likely they are to meet and interact. Communities thus serve as organizing principles of nodes in social networks and are created on shared affiliation, role, activity, social circle, interest or function.

\begin{figure}[t]
\centering
  \centering
  \epsfig{file=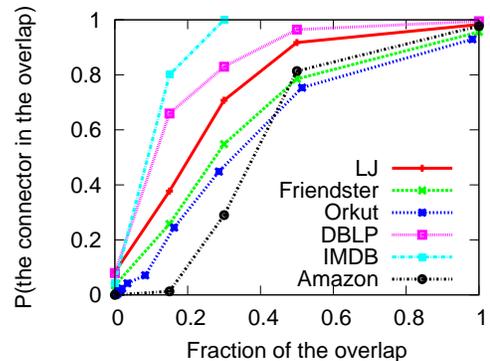, width=0.4\textwidth}
  \vspace{-4mm}
  \caption{Probability of a connector node belonging to the community overlap as a function of
  the fraction of community members in the overlap.}
  \label{fig:overlap.hub}
  \vspace{-4mm}
\end{figure}

\xhdr{Connector resides in the overlap}
%
The next question we investigate is whether the connector node (the node that is connected to the most other members of a community) belongs to the overlap. We extract all pairs of communities $(A,B)$ that have non-empty overlap $O$, and then compute the probability that the connector of community $A$ is in the overlap $O$. Figure~\ref{fig:overlap.hub} shows this probability as a function of the fraction of nodes in the community overlap ($|O|/|A|$). If the connector node would be a random node in a community, then the probability of a connector node belonging to the overlap is equal to the fraction of the nodes in the overlap, \ie, the probability of connector being in $O$ is $|O|/|A|$. However, we find that the probability of connector being in the overlap increases super-linearly with the size of the overlap in all the data sets. This demonstrates that connector nodes tend to reside in group overlaps and are not central to a single community.

Based on all these results, we believe that the dense community overlaps in our study reflect a fundamental property of the underlying networks. Understanding the possible causes for this property will be the subject of the next section.

\section{Implications of dense community overlaps}

\begin{figure}[t]
  \centering
  \begin{tabular}{ccc}
  \includegraphics[width=0.18\textwidth]{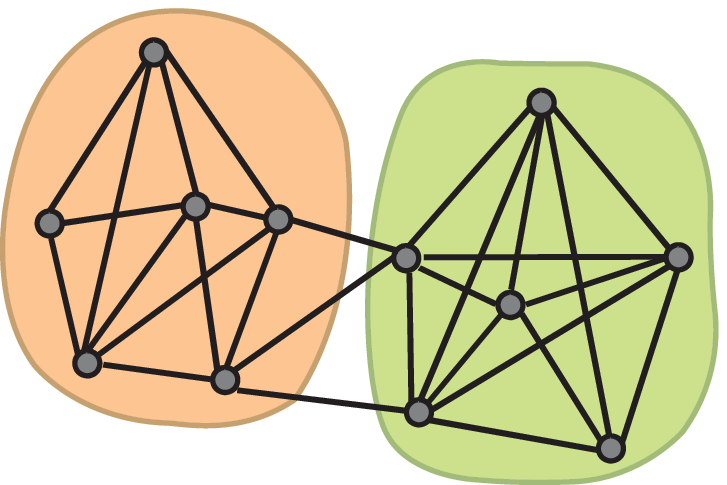} & &
  \includegraphics[width=0.18\textwidth]{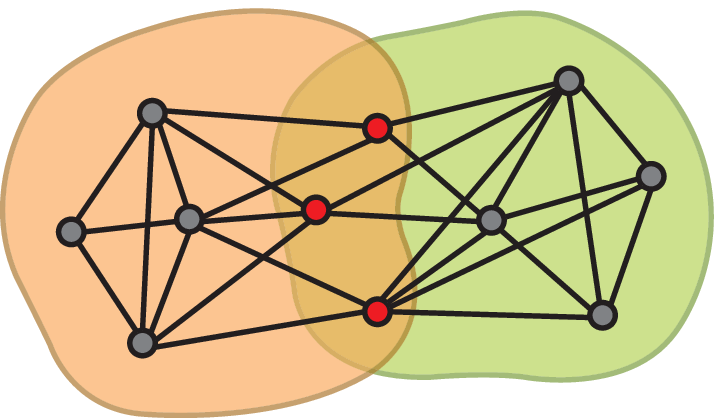} \\
  \subfigure[No overlaps]{\label{fig:introNonOver}\includegraphics[width=0.15\textwidth]{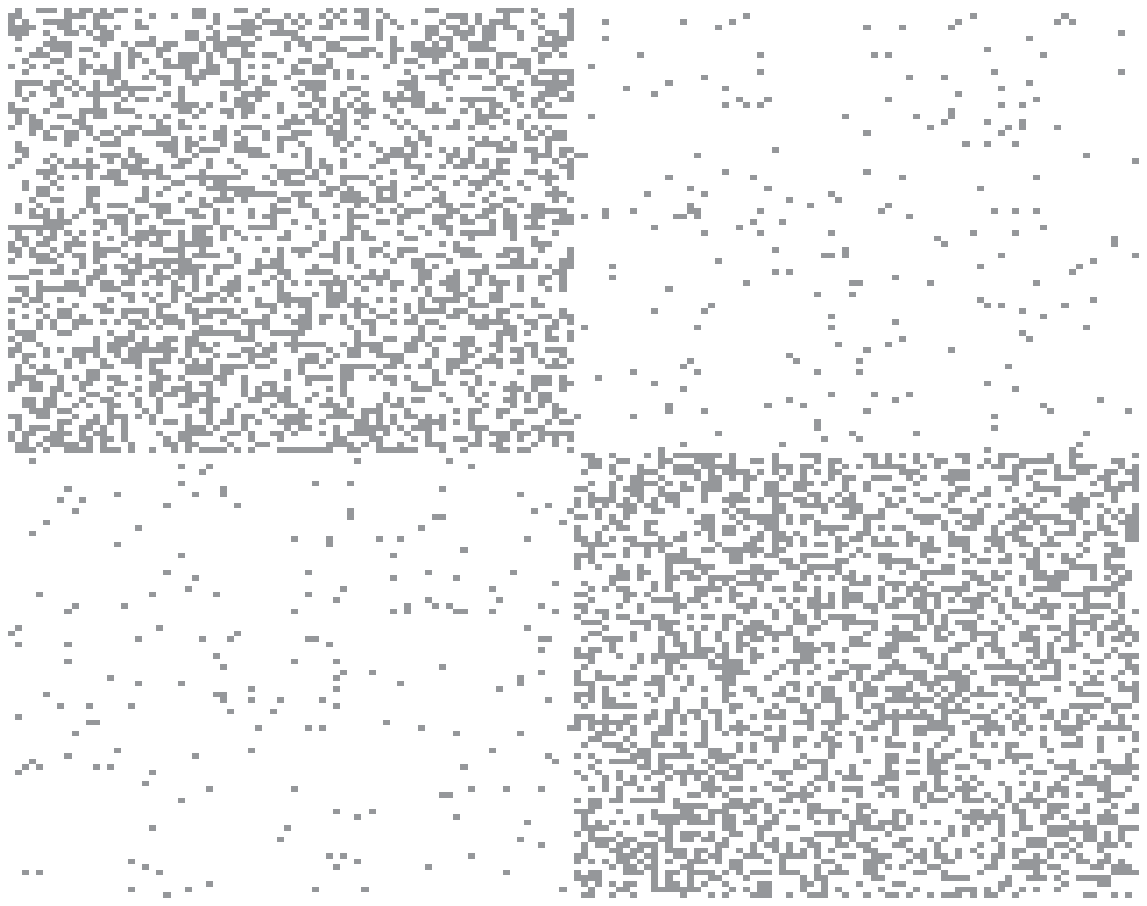}} & &
  \subfigure[Sparse overlaps]{\label{fig:introOver}\includegraphics[width=0.15\textwidth]{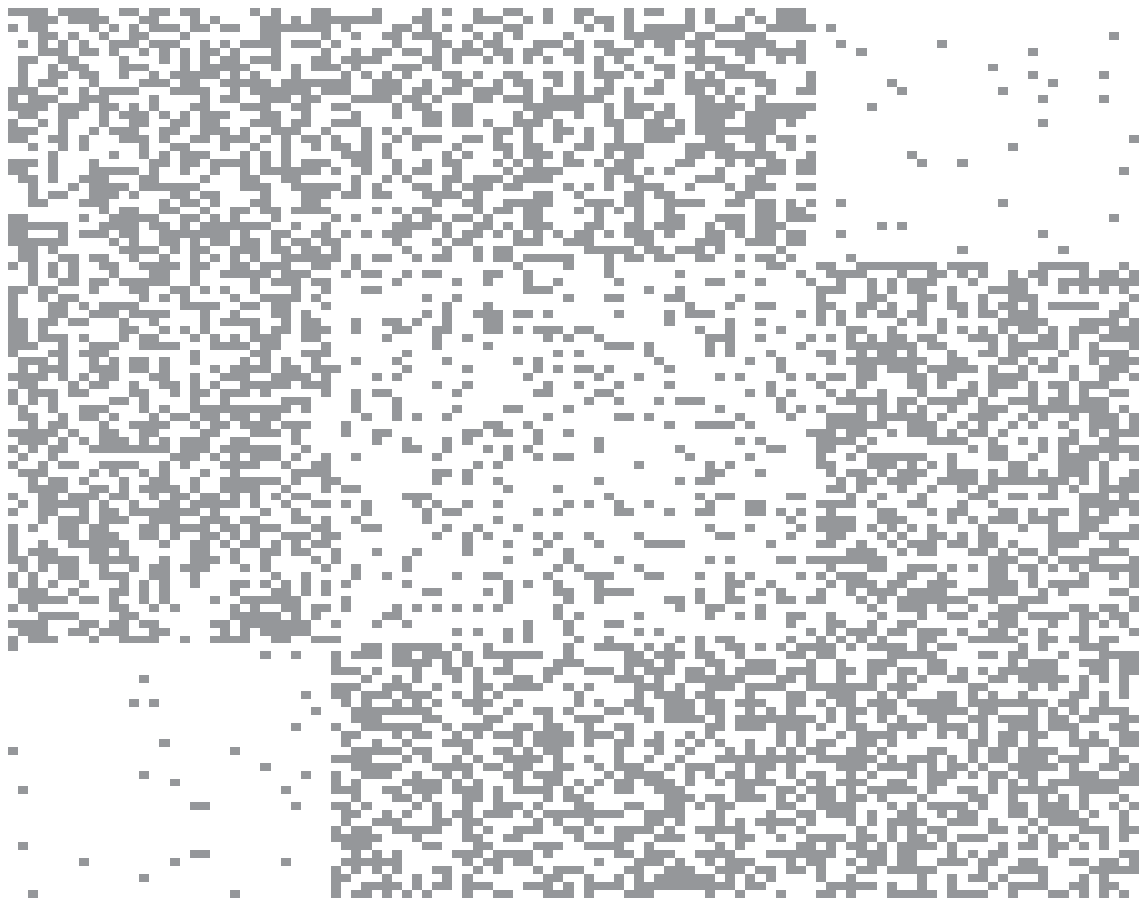}}
  \end{tabular}
  \subfigure[Dense overlaps]{\label{fig:agm.over} \includegraphics[width=0.18\textwidth]{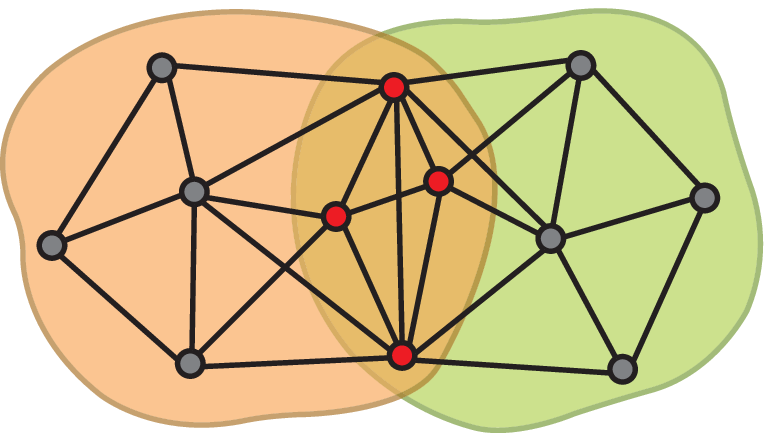}}
  \hspace{7mm}
  \subfigure[Adjacency mtx.]{\label{fig:agm.adj} \includegraphics[width=0.15\textwidth]{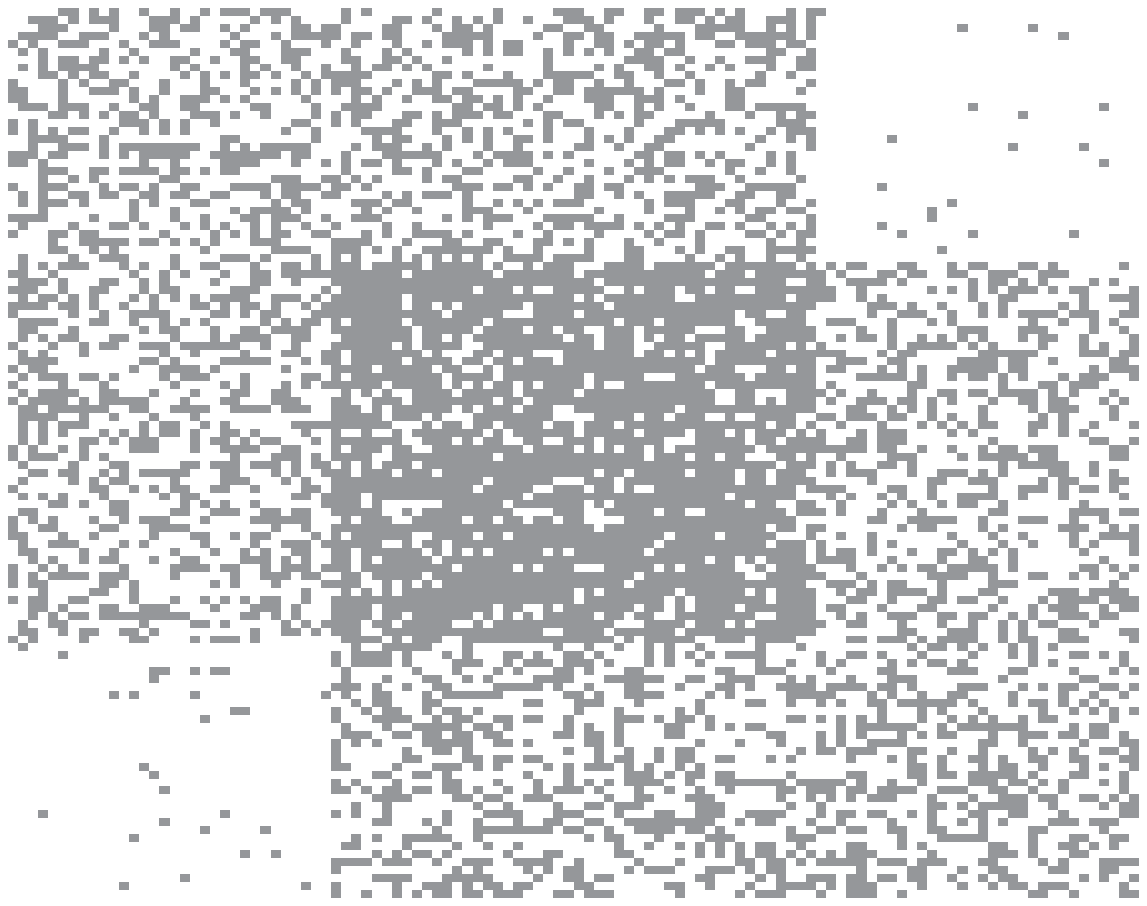}}
  \label{fig:intro}
  \label{fig:adjacencies}
  \vspace{-1mm}
  \caption{Conventional view of (a) two non-overlapping, and (b) overlapping communities.
  Top: network, bottom: corresponding adjacency matrix.
  (c, d) Community overlaps as suggested by our findings.}
  \vspace{-2mm}
\end{figure}

Even though our findings above are intuitively natural, we note a sharp contrast between the current understanding of network communities and our findings. Present view of network communities is based on two fundamental social network theories: triadic closure \cite{watts98collective} and `strength of weak ties' \cite{granovetter73ties}. Building on these two theories leads to a picture of network communities as illustrated in Figure \ref{fig:introNonOver}: Nodes inside communities link densely to one another while there are relatively few edges between the groups. Early works on network community detection, \eg, Newman's betweenness centrality~\cite{fortunato09community}, Modularity optimization~\cite{fortunato09community} as well as graph partitioning methods all adopt this view of network communities. This view of communities has another important consequence. It suggests that homophily in networks operates in small pockets where nodes gather in dense non-overlapping clusters (Fig. \ref{fig:introNonOver}).

In networks communities also tend to overlap as nodes can belong to multiple communities at once (and thus residing in the overlap) \cite{palla05_OveralpNature}. Applying the conventional view of network communities in this case leads to the (unnatural) structure of community overlaps as illustrated in Figure~\ref{fig:introOver}. Present models of overlapping communities \cite{Ahn10LinkCommunitiesNature,lancichinetti09ovlpbenchmark,palla05_OveralpNature} assume that community overlaps are {\em less densely} connected than the groups themselves. This means that they assume that the probability of an edge between a pair of nodes {\em decreases} with the number of shared community memberships. As a consequence this means that present methods cannot properly detect communities in large networks -- they would either mistakenly identify the overlap as a separate community or merge two overlapping communities into a single community.

Our finding that, the more community affiliations a pair of nodes shares, the more likely they are connected, suggests community overlaps as illustrated in Figure \ref{fig:agm.over}. This view of network formation differs from what has been assumed in the past and is consistent with early works in social network analysis. In particular, works of Simmel on the web of affiliations~\cite{simmel64affiliations}, and Feld on the focused organization of social ties~\cite{feld86focused} view networks as being composed of overlapping tiles or social circles that serve as organizing principles of nodes in networks. Our work suggests exactly the same analogy: Network community can be thought of as overlapping tiles and areas of the network where more tiles overlap naturally contain more connections.

With respect to homophily, our work extends its notion. In a sense we are discovering {\em pluralistic homophily} where the similarity of one node to another is the number of shared affiliations, not just their similarity along a single dimension. This means that homophily does not operate in concentrated pockets but rather as overlapping tiles. Last, the fact that regions of the network where communities overlap are more densely connected also nicely explains the co-existence of communities and the global core-periphery structure that has been observed in many networks~\cite{jure08ncp2}.

\section{Community-Affiliation\\ Graph Model}
\label{sec:model}

In the following, we would like to find some simple, conceptual model of behavior, which could naturally lead to the phenomena that we have observed. Building on Breiger's foundational work \cite{breiger74groups} where it has been recognized that communities and ``cliques'' arise due to shared group affiliations \cite{breiger74groups,simmel64affiliations,feld86focused}, we present the {\em \agmlong} (\agm), a family of simple probabilistic generative models for graphs that capture the observed phenomena and reliably reproduce the organization of networks into communities and the overlapping community structure.

\xhdr{\agmlong}
We build our model on the following intuition. Consider a pair of people that are members of several different interest based communities. Then, by having more interests in common, they are more likely to meet and link. Our model is thus based on two main ingredients. First ingredient is a bipartite affiliation network that links nodes of the social network to communities that they belong to. The second ingredient is the insight that each community also carries a single parameter that captures the probability that nodes belonging to that community to share a link. Thus, naturally, the more communities a pair of nodes shares, the higher is the probability of linking. Figure~\ref{fig:AGM.bipartite} illustrates the essence of our model. We start with a bipartite graph where the nodes at the bottom represent the nodes of the social network and the nodes on the top represent communities. The edges between nodes of the social network and the communities indicate community memberships. We denote the bipartite affiliation network as $B(V,C,M)$, where $V$ denotes the set of nodes of the original social network $G$, $C$ is a set of communities, and there is an edge $(u,c) \in M$ from node $u \in V$ to community $c \in C$ if node $u$ belongs to community $c$.


Now, given the affiliation network $B(V,C,M)$, we want to generate a social network graph $G(V,E)$. To achieve this we need to specify the process which generates the edges $E$ of $G$ given the affiliation network $B$. We consider a simple parameterization where we assign a parameter $p_c$ to every community $c \in C$. Parameter $p_c$ models the probability of an edge between two members of community $c$. In other words, we simply generate an edge between a pair of nodes that belong to community $c$ with probability $p_c$.
Each community $c$ creates edges independently. However, if the two nodes have already been connected via some other common community membership, then the duplicate edge is not included in the graph $G(V,E)$.

\begin{definition} Let $B(V,C,M)$ be a bipartite graph where $V$ is a set of nodes, $C$ is a set of communities, and an edge $(u,c) \in M$ connects node $u \in V$ to community $c \in C$ if $u$ belongs to community $c$. Also, let $\{p_c\}$ be a set of probabilities for all $c \in C$. Given affiliation network $B(V,C,M)$ and $\{p_c\}$, the \agmlong generates a graph $G(V,E)$ with the node set  $V$ and the edge set $E$ as follows. For each pair of nodes $u, v \in V$, the \agm creates edge $(u,v) \in E$ with probability $p(u,v)$:
\beq
p(u,v)= 1 - \prod_{k \in C_{uv}} (1- p_k),
\eeq
\label{eq:puv}
where $C_{uv} \subset C$ is a set of communities that $u$ and $v$ share ($C_{uv} = \{c | (u,c), (v,c) \in M\}$).
\label{def:agm}
\end{definition}

\begin{figure*}[t]
  \centering
  \begin{tabular}{ccccc}
  \includegraphics[width=0.25\textwidth]{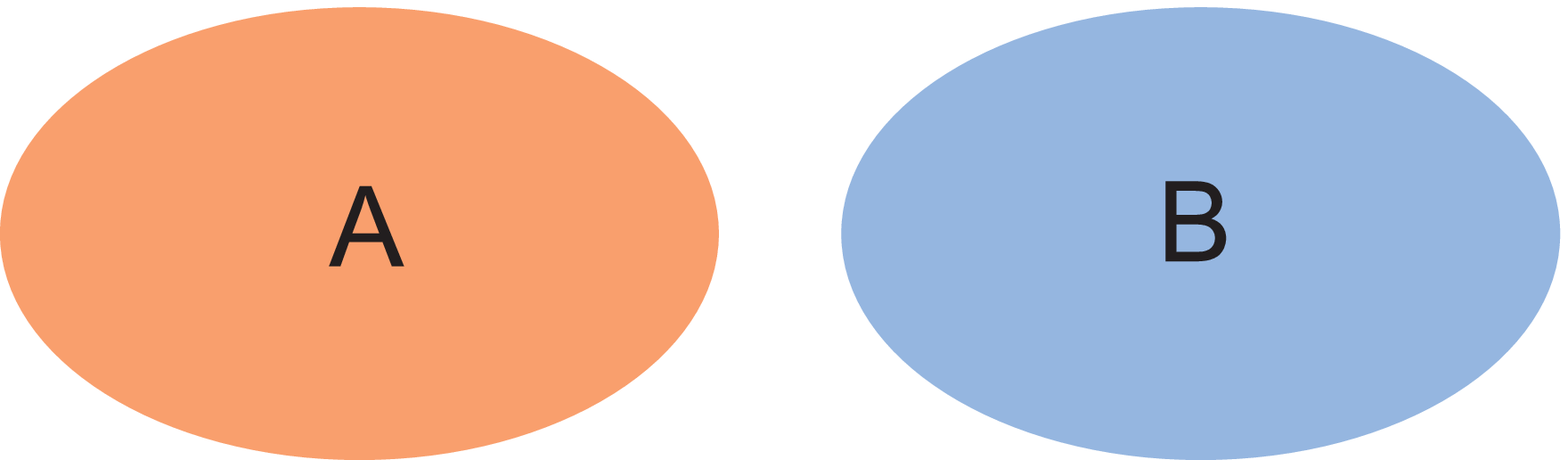} & &
  \includegraphics[width=0.25\textwidth]{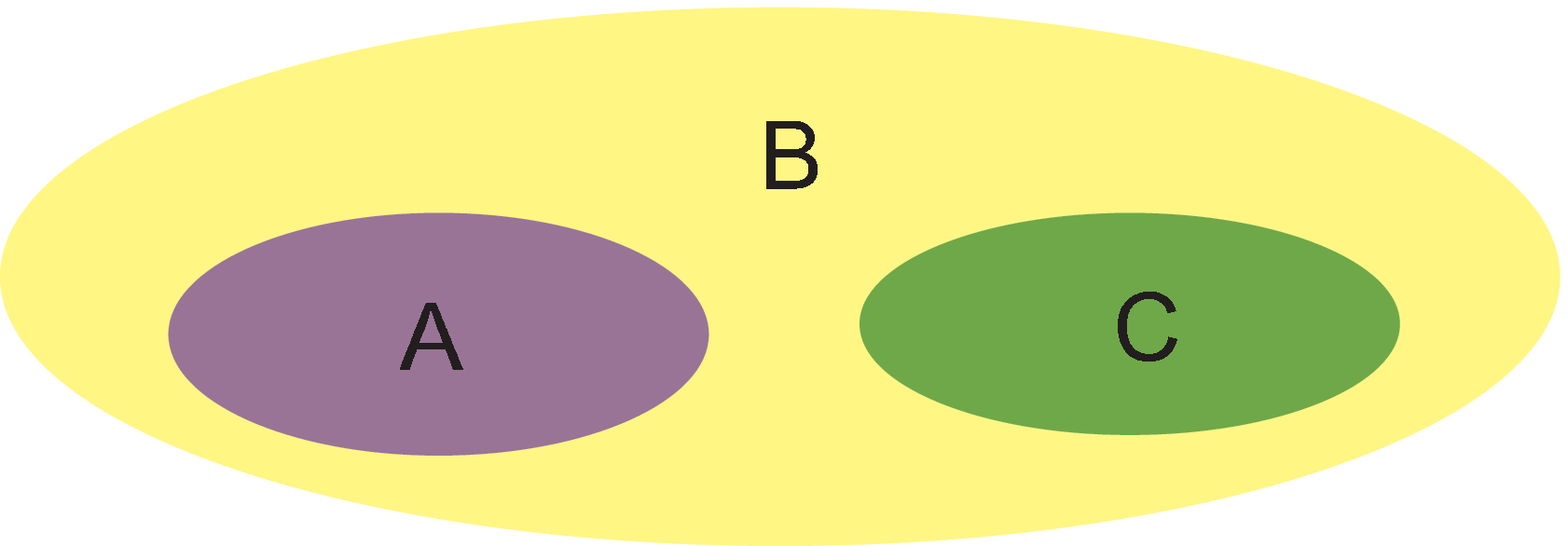} & &
  \includegraphics[width=0.23\textwidth]{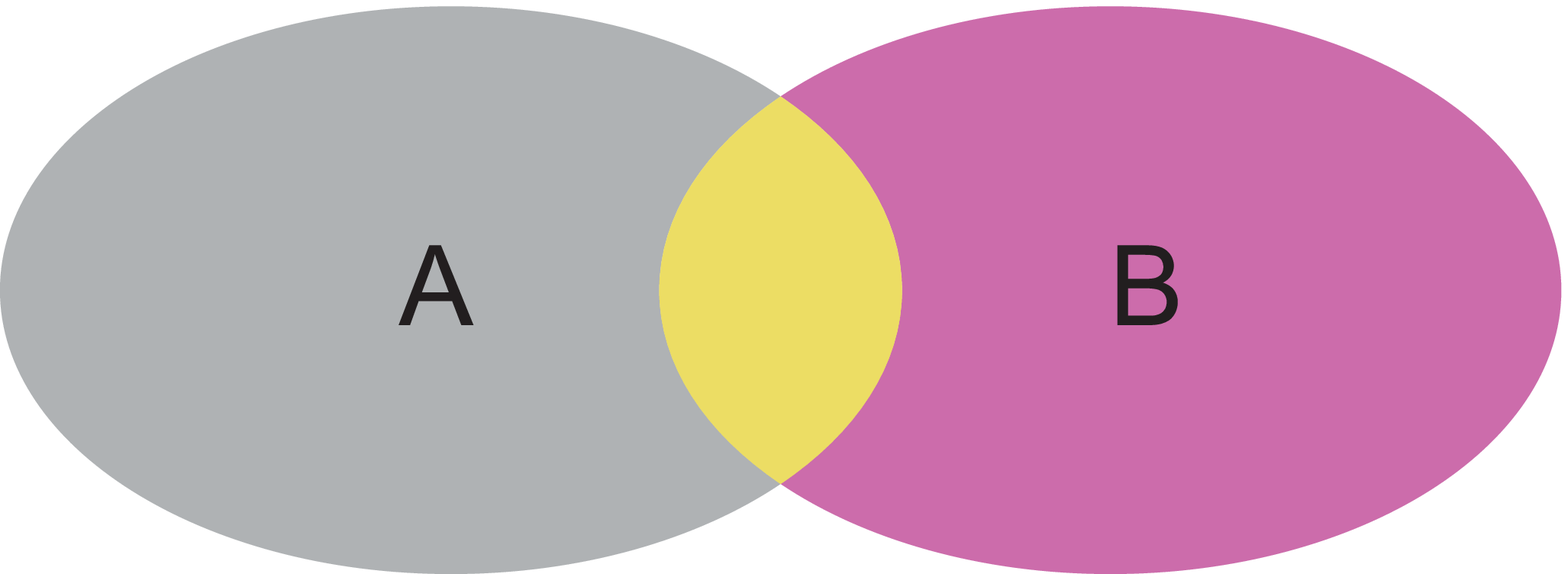}  \\
  \includegraphics[width=0.23\textwidth]{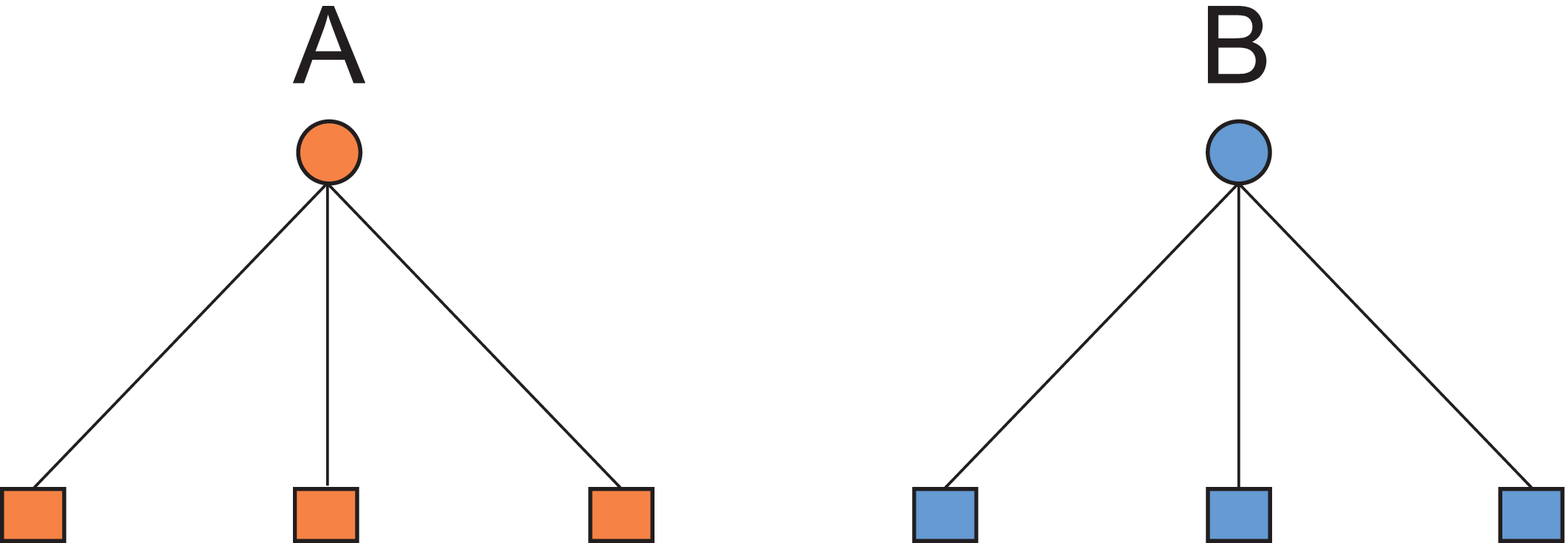} & &
  \includegraphics[width=0.33\textwidth]{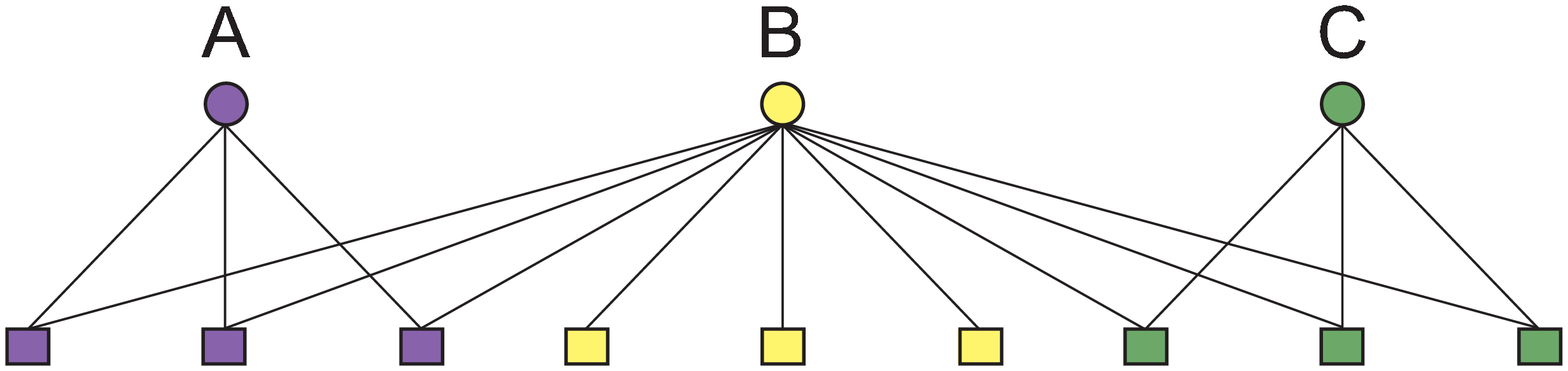} & &
  \includegraphics[width=0.23\textwidth]{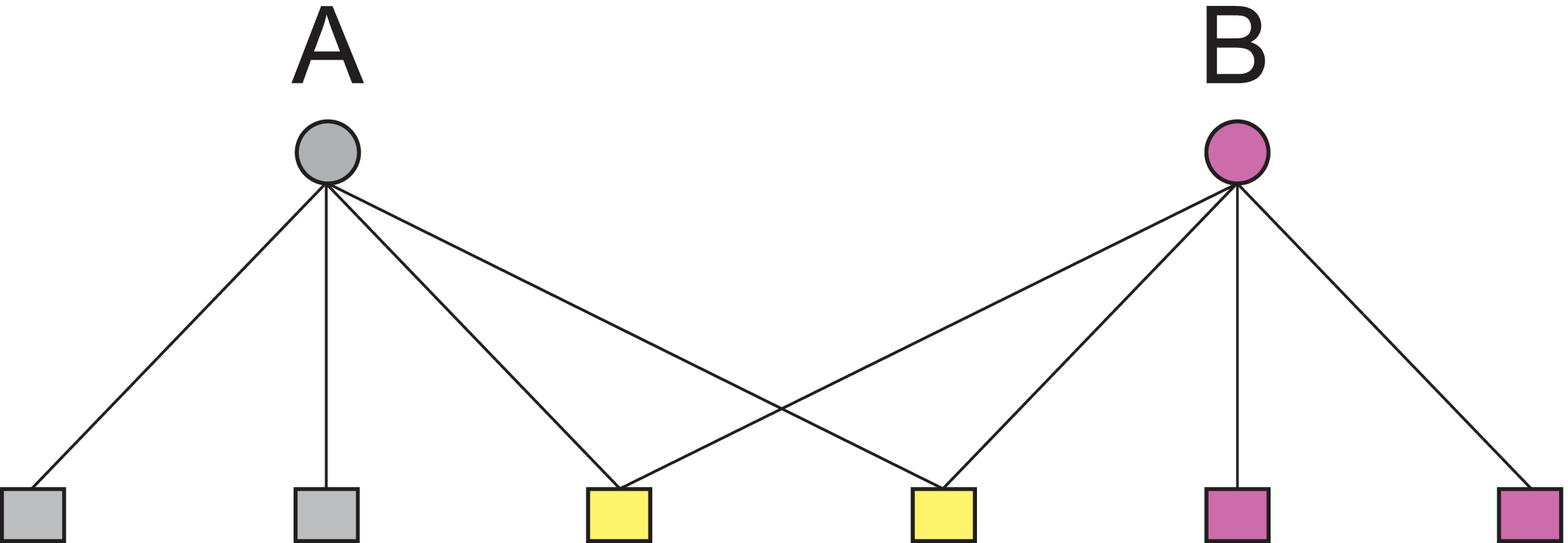} \\
  (a) Non-overlapping & &  (b) Nested & & (c) Overlapping
  \end{tabular}
  \label{fig:agm.flex}
  \caption{\agm allows for rich modeling of network communities: (a) non-overlapping, (b) nested, (c) overlapping. In (a) we can assume that nodes in disjoint communities connect with small prob. $\varepsilon$ which allows for sparse links between communities $A$ and $B$.
  }
\end{figure*}

Note that this simple process already ensures that pairs of nodes that belong to multiple common communities are more likely to link. This is due to the fact that nodes that share multiple community memberships get multiple chances to create a link. For example, pairs of nodes in the overlap of communities $A$ and $B$ in Figure~\ref{fig:AGM.bipartite} get two chances to create an edge. First they can get connected with probability $p_A$ (due to their membership in community $A$) and then also with probability $p_B$ (due to membership in $B$). While pairs of nodes residing in the non-overlapping region of $A$ link with probability $p_A$, nodes in the overlap link with probability $1-(1-p_A)(1-p_B) = p_A + p_B - p_A p_B \ge p_A$, which already ensures that overlaps of communities are more densely connected than the non-overlapping parts.

Last, we also point out the flexible nature of the \agmlong, which allows for modeling a wide range of network community structures. The flexibility of the affiliation network structure allows for modeling non-overlapping, hierarchically nested as well as overlapping communities (Figure~\ref{fig:agm.flex}).

\xhdr{Properties of the \agm}
The elegant nature of our models allows for mathematical analysis. Next, we derive several properties of the \agm networks that match the observations from Section~\ref{sec:experiments}. The aim of the analysis is to provide simple analytical results that illustrate that the \agm naturally obeys the empirical observations.


%
\begin{proposition}
    The expected number of edges $|E_c|$ between the nodes of community $c$ increases super-linearly as a function of the number of the nodes $n_c$ that belong to $c$, if $p_c$ is set to be proportional to $n_c^{-\beta}$ with $1 > \beta > 0$. (Observation in Figure~\ref{fig:sz.vol}).
    \label{pro:gdpl}
\end{proposition}
\begin{proofsketch}
    $|E_c| = \frac{ n_c (n_c - 1) }{2} \cdot p_c \propto n_c^2 p_c$.
    As $p_c \propto n_c^{-\beta}$ in the simplified \agm, we have $|E_c| \propto n_c^{2-\beta}$. As $\beta < 1$, $|E_c|$ grows super-linearly.
\end{proofsketch}

\begin{proposition}
    The fraction of connected neighbors in the overlap of two communities is higher than in the non-overlapping part of a single community. 
\end{proposition}
\begin{proofsketch}
    The fact naturally follows from Definition~\ref{def:agm}.
\end{proofsketch}

\begin{proposition}
    Given communities $A, B$ and their overlap $O$, the probability that the connector node (the node that is connected to most other members of a community) of $A$ is in overlap $O$ is higher than $|O| / |A|$ (Observation in Figure~\ref{fig:overlap.hub}).
\end{proposition}
\begin{proofsketch}
\agm generates edges among the nodes in community $A$ (node set $n_A$) independently with probability $p_A$ ($G_{n_A,p_A}$), and edges among the nodes in $B$ independently with prob. $p_B$ ($G_{n_B,p_B}$). Let $X_o$ be the event that a particular node $o \in O$ is the connector of $A$.
The probability of the connector node being in $O$ is the sum of $X_o$ for each $o \in O$, $\sum_{o \in O} p(X_o)$,
which is $ |O| \cdot p(X_o)$ as $p(X_o)$ is the same for any $o \in O$.
Now, it suffices to show that $p(X_o) > 1/|A|$.
$X_o$ is equivalent to the event that the internal degree of $o$ and $A$, $d_{in}(o,A)$, is the maximal among $d_{in}(u, A)$ of all $u \in A$.
$o$ is connected to other nodes in $A$ by both $G_{n_A,p_A}$ and $G_{n_B,p_B}$ as $G_{n_B,p_B}$ connects $o$ to other nodes in $O \subset A$.
Only if $G_{n_B,p_B}$ does not make any connection between $o$ and other members in $O$,
$d_{in}(o,A)$ has the same distribution as $d_{in}(u,A)$ of any other $u \in A$, and thus $p(X_o) = 1/|A|$.
However, $G_{n_B,p_B}$ connects $o$ to other members in $O$ with positive probability, and thus $p(X_o)$ is strictly higher than $1/|A|$.
\end{proofsketch}

\begin{proposition}
    If $p_c = p$ for all communities $c$, the conditional edge probability between two nodes is an increasing function of the number of communities that the both nodes belong to (Observation in Figure~\ref{fig:edge_prob}).
\end{proposition}
\begin{proofsketch}
    When two nodes belong to $k$ common communities, the \agm connects the two nodes with probability $1 - (1 - p)^k$, which is an increasing function of $k$.
\end{proofsketch}

\section{Model evaluation}
\label{sec:eval}
Having defined the \agm, we now proceed to investigate its properties.  We perform a set of simulation experiments and compare our model to other models of network community structure. For each network, we generate a synthetic network with the \agm, and then compare the synthetic network and the synthetic community structure to the structure of the real network. For a comparison, we use the model of network community structure proposed by Lancichinetti et al.~\cite{lancichinetti09ovlpbenchmark}. We refer to the model as the LFR. The LFR model is the state-of-the-art model for generating networks with overlapping community structure that can then be used for evaluating community detection methods. Our goal is to understand whether the \agm qualitatively reproduces structural properties of real networks, real communities and real community overlaps.



\xhdr{Experimental setup}
In order to compare real networks to the synthetic networks generated by the \agm and LFR, we need to set parameters of both models. Both the \agm and the LFR require bipartite affiliation networks, and for simplicity we construct the community affiliation network from the node community membership information. Then, we use maximum likelihood estimation to fit the parameters of the LFR as well as the \agm.
For \agm, we fit a set of probabilities $\{p_c\}$. (We discuss the fitting of $\{p_c\}$ via convex optimization in the next subsection.) The LFR requires the following parameters: the power-law coefficient of the network degree distribution and the fraction of external edges of each node. We estimate these parameters from the real network using maximum likelihood estimation. Note that when computing the community internal degree of a node, LFR penalizes nodes in the overlap: the internal degree is inversely proportional to the number of communities that the node belongs to. Therefore, LFR also assumes sparse community overlaps.


We then compare the networks synthesized by the two models to the ground-truth networks. We investigate three criteria: structure of communities, structure of community overlaps, and structure of the networks themselves. For each of the six datasets we repeat the measurements from Section~\ref{sec:experiments} on real and on synthetic networks and examine the performance of the two models. For brevity we focus on LiveJournal and refer the reader to the extended version of the paper~\cite{jaewon11community} for results on other datasets.

\xhdr{Estimating $p_c$ via convex optimization}
Given a graph $G(V,E)$ and a bipartite community affiliation network $B(V, C, M)$, we aim to find parameters $\{p_c\}$ that maximize the likelihood of observed edges in $G$:

\begin{equation}
L(\{p_c\})  = \prod_{(u, v) \in E} p(u,v) \prod_{(u, v) \not\in E} (1-p(u,v)).
\label{eq:ll}
\end{equation}

By applying Eq.~\ref{eq:puv} we transform the optimization problem to:
\[
\arg \max_{\{p_c\}} \sum_{(u, v) \in E} (1 - \prod_{k \in C_{uv}} (1- p_k)) \sum_{(u, v) \not\in E} (\sum_{k \in C_{uv}} (1- p_k))
\]
with constraints $0 \leq p_c \leq 1$.

This optimization is nontrivial to solve. The objective function is non-convex as it involves a product over the variables $p_k$. Now, we show that it can be converted to a convex optimization problem.

We maximize the logarithm of the likelihood and perform a change of variables $1 - p_k = e^{-x_k}$:
\[
\arg \max_{\{x_c\}} \sum_{(u, v) \in E} \log (1 - e^ { - \sum_{k \in C_{uv}} x_k}) - \sum_{(u, v) \not\in E} \sum_{k \in C_{uv}}  x_k,
\]
and constraints $0 \leq p_c \leq 1$ become $x_c \geq 0$. This transformed problem is a convex function of $\{x_c\}$ and thus the globally optimal optimal values of $\{x_c\}$ can be efficiently found. Then, by the change of variables, we find the values of $\{p_c\}$.

\begin{figure}[t]
\centering
  \centering
  \subfigure[Edges Inside the group]{\label{fig:sz.vol.models}\includegraphics[width=0.23\textwidth]{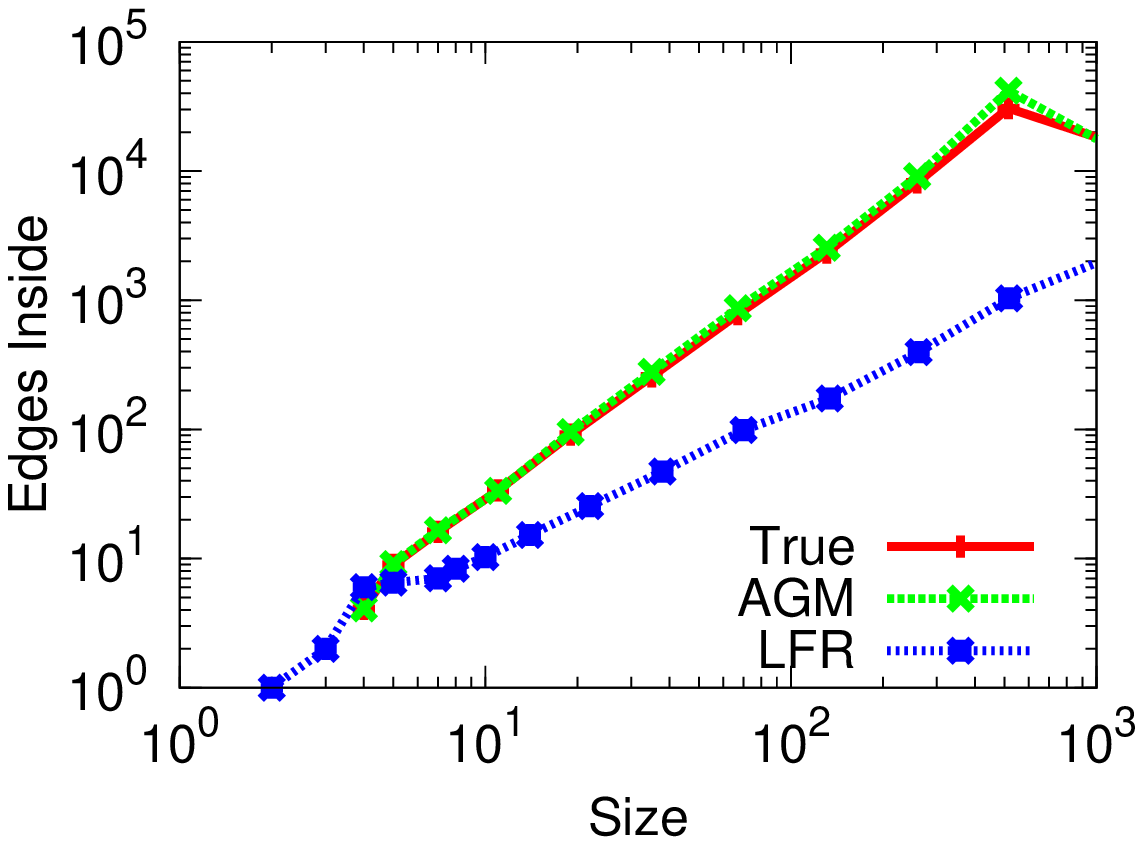}}
  \subfigure[Maximal ICDF]{\label{fig:sz.maxintdeg.models}\includegraphics[width=0.23\textwidth]{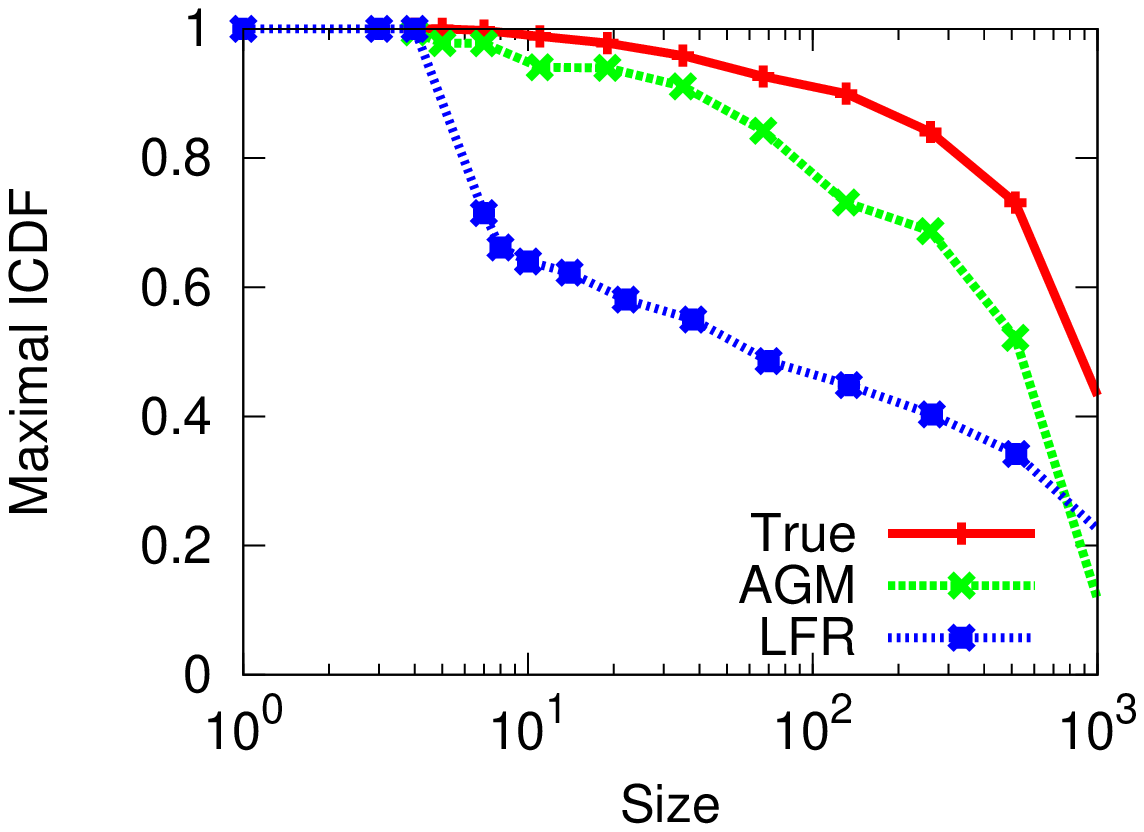}}
  \caption{LiveJournal community properties.}
  \label{fig:group.stat.models}
\vspace{-5mm}
\end{figure}

\xhdr{Evaluation: Properties of communities}
First, we compare the connectivity patterns of communities in the synthetic networks to the ground-truth LiveJournal network. We perform measurements analogous to those in Figure~\ref{fig:group.stat} and plot the results in Figure~\ref{fig:group.stat.models}. We overlay the original results from LiveJournal (Figure~\ref{fig:group.stat}) with the properties of communities in the synthetic \agm and LFR networks. We observe that the \agm much better captures connectivity patterns of communities. For example, communities in the \agm tend to have connector nodes when the community size is smaller than $100$ nodes (Figure~\ref{fig:sz.maxintdeg.models}). Similarly, the \agm also captures the community densification power-law nearly perfectly (Figure~\ref{fig:sz.vol.models}).

\xhdr{Evaluation: Community overlaps}
In Section~\ref{sec:experiments}, we observed that nodes in the overlap (multi membership nodes) have higher connectivity than the nodes not in the overlap (single membership nodes). We also noted that an overlap is likely to contain the connector node of the group, and the overlap becomes denser as more groups join the overlap. We examine how well the two models mimic these patterns of group overlaps.

\begin{figure}[t]
\centering
  \subfigure[Edge probability]{\label{fig:edge_prob.models}\includegraphics[width=0.23\textwidth]{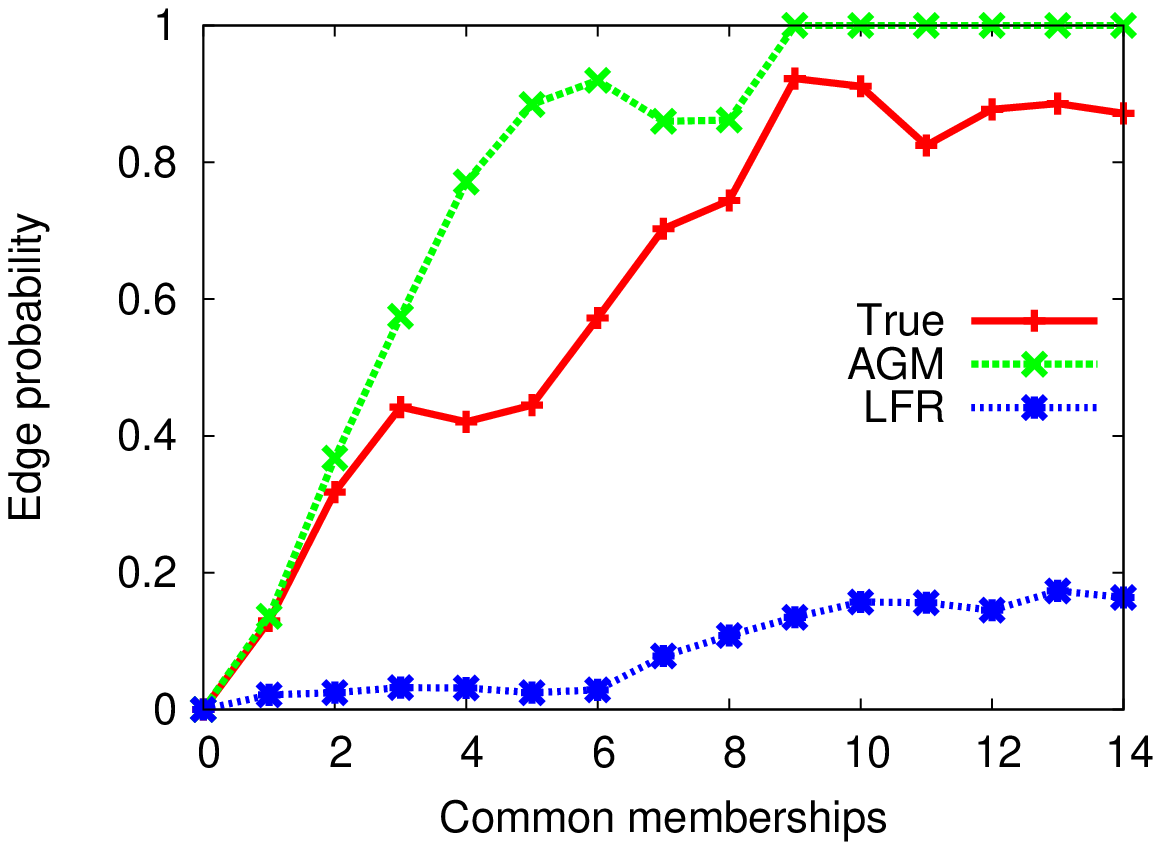}}
  \subfigure[Connector inside the overlap]{\label{fig:hub.overlap.models}\includegraphics[width=0.23\textwidth]{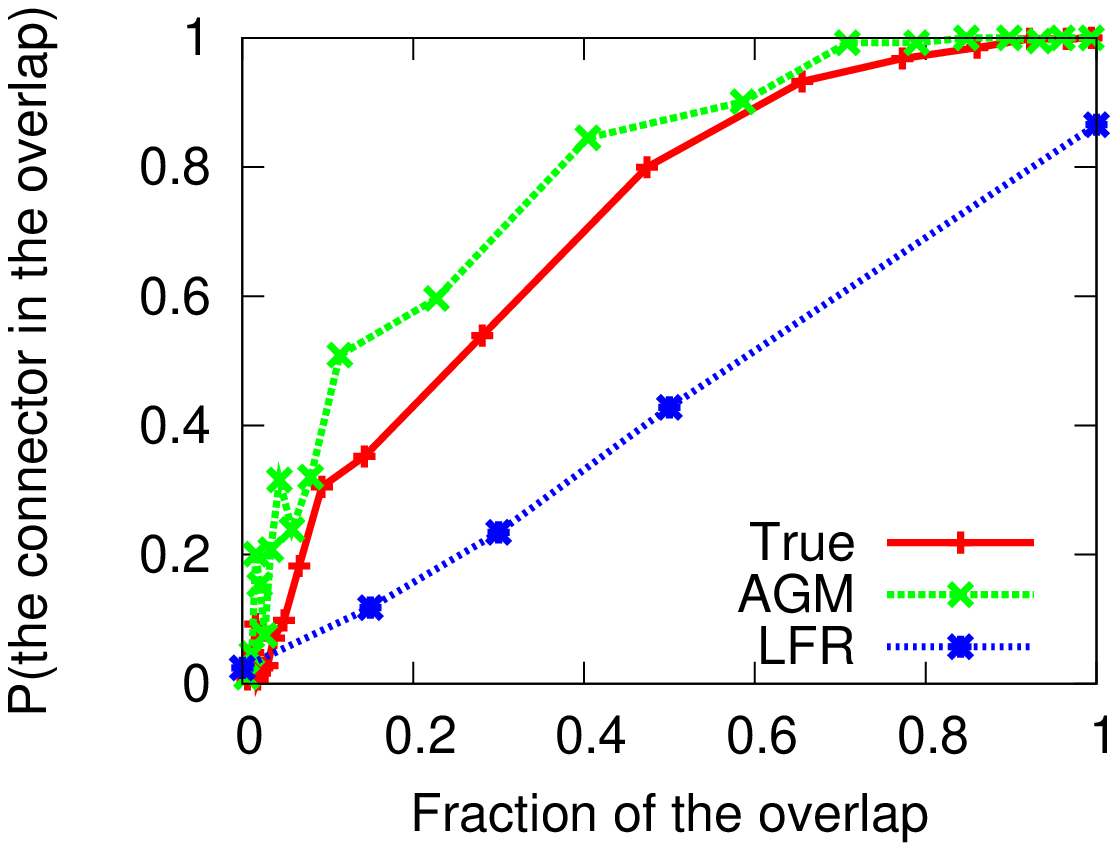}}
\vspace{-2mm}  
  \caption{LiveJournal community overlaps.}
\vspace{-5mm}  
\end{figure}

First, we found that the edge probability between a pair of nodes is an increasing function of the number of communities that the nodes share (Figure~\ref{fig:edge_prob}). Figure~\ref{fig:edge_prob.models} plots the edge probability as a function of the number of common communities between a pair of nodes for the LiveJournal network and compares it to the two models. Notice that the \agm successfully reproduces the edge probability, while LFR fails to model the fact that nodes that share more communities tend to be more likely to be connected. In fact, Figure~\ref{fig:edge_prob} shows the edge probability for all six datasets (red lines) and the same probability as modeled by the \agm (green lines). Notice that the \agm is able to capture a wide range of behaviors --- from diminishing-returns (Fig. \ref{fig:edge_prob.lj}), S-shape (Fig. \ref{fig:edge_prob.dblp}) to a slowly rising pattern (Fig. \ref{fig:edge_prob.amazon})


Second, we also observed that real communities have a connector node and the connector node is more likely to exist inside the overlap (Figure~\ref{fig:overlap.hub}). We validate the presence of the connector node in synthetic communities in Figure~ \ref{fig:hub.overlap.models}. We compute the probability that the overlap $O$ between two communities $A$ and $B$ has a connector node of either community $A$ or $B$. 
Notice a very close fit to the LiveJournal network. In contrast, in LFR, the probability of a connector being in the overlap is much lower, which confirms that the overlaps between LFR communities are less dense than a single community.

This is interesting as it explains why communities in the \agm tend to have connector nodes (Figure~\ref{fig:sz.maxintdeg.models}). Since edges inside each \agm community are created independently (which by itself does not produce skewed node degrees) we would naturally expect that all the nodes would have similar degrees and no connector nodes would emerge. However, since the overlaps are dense, the nodes in the overlap tend to have higher degrees and emerge as connector nodes (Fig.~\ref{fig:sz.maxintdeg.models}). LFR explicitly forces a heavy-tailed degree distribution which makes a few nodes have very high degrees. However, LFR communities do not have a connector node. This is because LFR prevents the nodes in the overlaps from forming edges in a single community and thus the connector node links to a small fraction of community members.

We now briefly mention the performance of the \agm on all other datasets. For each structural property, we measure the quality of fit between that property in the synthetic \agm and LFR networks and the real data. We apply the Kolmogorov-Smirnov (KS) statistic, which is a non-parametric way of quantifying the distance between two distribution functions. Given two distribution functions $f(x), g(x)$, \ie, plots of the same structural property, the KS-statistic computes the maximum difference between the cumulative area under the two curves, $KS(f,g) = \sup_x |\int^x f(t) dt  - \int^x g(t) dt|$. We compute the KS-statistic between the AGM curve (or the LFR curve) and the true curve for all the following properties:
\begin{itemize}
\denselist
    \item \emph{VOL:} Edges inside the community. 
    \item \emph{MID:} Maximum Internal Community Degree Fraction. 
    \item \emph{EP:} Edge probability between nodes. 
    \item \emph{PC:} Probability of a connector residing in the overlap. 
    \item \emph{OO:} Fraction of connected neighbors in the overlap. 
    \item \emph{AABB:} Fraction of connected neighbors in a community. 
\end{itemize}

\begin{table}[t]
  \centering
  \small
  \begin{tabular}{l||r|r|r|r|r|r||r}
        Network & Vol & MID & PC & EP & OO & AABB & Avg\\ \hline \hline
    LiveJournal&0.98&0.72&0.53&0.73&0.86&0.77&0.76\\ \hline
    Friendster&1.00&0.55&0.70&0.84&0.63&0.49&0.70\\ \hline
    Orkut&0.99&0.67&0.57&0.95&0.50&0.21&0.65\\ \hline
    DBLP&0.99&0.05&0.88&0.66&0.66&0.56&0.63\\ \hline
    IMDB&0.93&0.36&0.17&0.39&0.16&0.21&0.37\\ \hline
    Amazon&0.91&0.73&0.35&-0.79&0.80&0.76&0.46\\ \hline \hline
    Average&0.97&0.51&0.53&0.46&0.60&0.50&0.60
  \end{tabular}
\vspace{-2mm}  
  \caption{Community connectivity and overlaps: Relative improvement in the KS-statistic of the \agm over LFR.}
  \label{table:ks.group}
\vspace{-3mm}  
\end{table}



Table~\ref{table:ks.group} reports the relative improvement in the KS-statistic between the two models (the difference between the two models normalized by the larger value of the two). The value of the relative improvement can be between 1 (the \agm completely outperforms LFR) and -1 (LFR completely outperforms the \agm).
The \agm shows a relative improvement of 60\%, which means that the \agm outperforms LFR by a factor of 2.
Furthermore, the \agm shows significantly lower average KS-statistics in every property as well as on every network. Overall, we conclude that the \agm reliably captures the properties of real overlaps and real groups and significantly improves over previous state of the art models.

\xhdr{Evaluation: Network properties}
Last, we also study whether the \agm is able to generate overall realistic networks. We examine how well the global structural properties of the synthetic networks match the properties of the ground-truth network. For each of the networks synthesized by the \agm and LFR, we quantify the degree of agreement between the real and synthetic network by computing the KS-statistic on the following network properties:
\begin{itemize}
\denselist
    \item \emph{Degree distribution (Deg):} histogram of the number of edges of a node. Networks tend to have power-law degrees~\cite{barabasi99emergence}.
    \item \emph{Clustering coefficient (CCF):} distribution of clustering coefficient of nodes~\cite{watts98collective}.
    \item \emph{Hop plot (Hop):} the number of reachable pairs of nodes in less than $x$ hops~\cite{jure05dpl}.
    \item \emph{Triad participation (TP):} the number of triangles that a node participates in~\cite{tsourakis08triangles}.
    \item \emph{Eigenvalues (EigVal):} distribution of eigenvalues of the adjacency matrix~\cite{chakrabarti06survey}.
    \item \emph{Eigenvector (EigVec):} distribution of components in the eigenvector associated with the largest eigenvalue~\cite{chakrabarti06survey}. 
\end{itemize}

%

Table \ref{table:ks.network} shows the relative improvements in KS-statistics of the \agm over LFR. The \agm network follows very closely the patterns of the ground-truth network for most properties, like the degree distribution, triad participation, and eigenvalues. The only exception where LFR outperforms the \agm is the Eigenvector. The \agm exhibits 9\% better fit (KS-statistic) than LFR for Degree distribution, 221\% for Clustering coefficient, 7\% for Hop distribution, 120\% for Triad participation, and 122\% for Eigenvalues. For the Eigenvector property, the LFR exhibits a 17\% better value. Overall, these results demonstrate that the \agm is not only able to reliably capture the structure of network communities and community overlaps but that it also accurately generates the underlying networks.


\begin{table}[t]
\centering
\small
  \begin{tabular}{c||c|c|c|c|c|c}
        Network&Deg&CCF&Hop&TP&EigVal&EigVec\\ \hline \hline
        LiveJournal&1.09&2.72&-0.03&1.72&2.52&-0.29\\ \hline
        Friendster&0.35&2.37&0.04&1.61&1.74&0.24\\ \hline
        Orkut&1.06&1.52&0.08&3.39&1.12&0.51\\ \hline
        DBLP&-1.10&2.31&0.02&1.07&0.59&-0.74\\ \hline
        IMDB&-0.03&1.86&0.20&0.69&1.87&-0.51\\ \hline
        Amazon&-0.81&2.48&0.12&-1.29&-0.53&-0.24\\ \hline\hline
        Average KS&0.09&2.21&0.07&1.20&1.22&-0.17
   \end{tabular}
  \vspace{-1mm}
  \caption{Relative difference in the KS-statistic of \agm and LFR for network properties. Positive values mean that \agm outperforms LFR.}
  \label{table:ks.network}
  \vspace{-2mm}
  \end{table}

\section{Conclusion}
\label{sec:conclusion}



In this paper we identified a set of networks with explicitly defined ground-truth communities. This allowed us to investigate the structure and overlaps of ground-truth communities in networks. We observed that the overlaps of communities are more densely connected than the non-overlapping parts of communities, which is in contrast to assumptions made by present community detection models and methods. We also observed that ground-truth communities contain high-degree hub nodes that reside in community overlaps and link to most of the members of the community. We then presented the {\em \agmlong} (\agm), a conceptual model of network community structure, which reliably captures the overall structure of networks as well as the overlapping nature of network communities.

Our results have relevance in multiple settings. First, our analysis sheds light on the organization of complex networks and provides new directions for research on community detection. Second, ground-truth communities offer a reliable ground-truth for community evaluation that was impossible to do before. Last, the \agm provides a realistic benchmark network on which new community detection algorithms can be developed and evaluated.

A natural step for future work is build on these findings and design community detection methods that can detect dense overlaps. Explicitly maximizing the likelihood defined in Eq. \ref{eq:ll} over both the affiliation graph $B$ as well as the parameters $p_c$ would be a good step in this direction.




%

\xhdr{Acknowledgements}
This research has been supported in part by NSF
CNS-1010921,         
IIS-1016909,            
IIS-1149837,            
IIS-1159679,            
Albert Yu \& Mary Bechmann Foundation, Boeing, Allyes, Samsung,
Alfred P. Sloan and the Microsoft Faculty Fellowship.


\begin{thebibliography}{10}

\bibitem{Ahn10LinkCommunitiesNature}
Y.-Y. Ahn, J.~P. Bagrow, and S.~Lehmann.
\newblock {Link communities reveal multi-scale complexity in networks}.
\newblock {\em Nature}, 2010.

\bibitem{barabasi99emergence}
R.~Z. Albert and A.-L. Barab\'{a}si.
\newblock Emergence of scaling in random networks.
\newblock {\em Science}, 286(5439):509--512, 1999.

\bibitem{lars06groups}
L.~Backstrom, D.~Huttenlocher, J.~Kleinberg, and X.~Lan.
\newblock Group formation in large social networks: membership, growth, and
  evolution.
\newblock In {\em KDD '06}, 2006.

\bibitem{barabasi04bionets}
A.-L. Barab\'{a}si and Z.~N. Oltvai.
\newblock {Network biology: understanding the cell's functional organization}.
\newblock {\em Nature Reviews Genetics}, 5(2):101--113, 2004.

\bibitem{breiger74groups}
R.~L. Breiger.
\newblock The duality of persons and groups.
\newblock {\em Social Forces}, 53(2):181--190, 1974.

\bibitem{chakrabarti06survey}
D.~Chakrabarti and C.~Faloutsos.
\newblock Graph mining: Laws, generators, and algorithms.
\newblock {\em ACM Computing Survey}, 2006.

\bibitem{feld86focused}
S.~L. Feld.
\newblock The focused organization of social ties.
\newblock {\em American Journal of Sociology}, 86(5):1015--1035, 1981.

\bibitem{flake00_efficient}
G.~Flake, S.~Lawrence, and C.~Giles.
\newblock Efficient identification of web communities.
\newblock In {\em KDD '00}, 2000.

\bibitem{fortunato09community}
S.~Fortunato.
\newblock Community detection in graphs.
\newblock {\em Physics Reports}, 486(3--5):75--174, 2010.


\bibitem{granovetter73ties}
M.~S. Granovetter.
\newblock The strength of weak ties.
\newblock {\em American Journal of Sociology}, 78:1360--1380, 1973.

\bibitem{kairam12ning}
S.~Kairam, D.~Wang, and J.~Leskovec.
\newblock The life and death of online groups: Predicting group growth and
  longevity.
\newblock In {\em WSDM '12}, 2012.


\bibitem{kumar00stochastic}
R.~Kumar, P.~Raghavan, S.~Rajagopalan, D.~Sivakumar, A.~Tomkins, and E.~Upfal.
\newblock Stochastic models for the web graph.
\newblock In {\em FOCS '00}, 2000.

\bibitem{lancichinetti09ovlpbenchmark}
A.~Lancichinetti and S.~Fortunato.
\newblock Benchmarks for testing community detection algorithms on directed and
  weighted graphs with overlapping communities.
\newblock {\em Phys. Rev. E}, 2009.

\bibitem{Lattanzi09AffiliationNetworks}
S.~Lattanzi and D.~Sivakumar.
\newblock Affiliation networks.
\newblock In {\em STOC '09}, 2009.

\bibitem{jure07viral}
J.~Leskovec, L.~Adamic, and B.~Huberman.
\newblock The dynamics of viral marketing.
\newblock {\em ACM TWEB}, 1(1), 2007.

\bibitem{jure05dpl}
J.~Leskovec, J.~Kleinberg, and C.~Faloutsos.
\newblock Graphs over time: densification laws, shrinking diameters and
  possible explanations.
\newblock In {\em KDD '05}, 2005.

\bibitem{jure07evolution}
J.~Leskovec, J.~Kleinberg, and C.~Faloutsos.
\newblock Graph evolution: Densification and shrinking diameters.
\newblock {\em ACM TKDD}, 2007.

\bibitem{jure08ncp2}
J.~Leskovec, K.~J. Lang, A.~Dasgupta, and M.~W. Mahoney.
\newblock Community structure in large networks: Natural cluster sizes and the
  absence of large well-defined clusters.
\newblock {\em Internet Mathematics}, 6, 2009.

\bibitem{mcpherson01homophily}
M.~McPherson, L.~Smith-Lovin, and J.~M. Cook.
\newblock Birds of a feather: Homophily in social networks.
\newblock {\em Annual Review of Sociology}, 27(1):415--444, 2001.

\bibitem{mislove07measurement}
A.~Mislove, M.~Marcon, K.~P. Gummadi, P.~Druschel, and B.~Bhattacharjee.
\newblock Measurement and analysis of online social networks.
\newblock In {\em IMC '07}, 2007.

\bibitem{mitzenmacher04brief}
M.~Mitzenmacher.
\newblock A brief history of generative models for power law and lognormal
  distributions.
\newblock {\em Internet Mathematics}, 1(2):226--251, 2004.

\bibitem{palla05_OveralpNature}
G.~Palla, I.~Der\'{e}nyi, I.~Farkas, and T.~Vicsek.
\newblock Uncovering the overlapping community structure of complex networks in
  nature and society.
\newblock {\em Nature}, 435(7043):814--818, 2005.

\bibitem{papadopoulos11community}
S.~Papadopoulos, Y.~Kompatsiaris, A.~Vakali, and P.~Spyridonos.
\newblock Community detection in social media.
\newblock {\em Data Mining and Knowledge Discovery}, 2011.

\bibitem{RCCLP04_PNAS}
F.~Radicchi, C.~Castellano, F.~Cecconi, V.~Loreto, and D.~Parisi.
\newblock Defining and identifying communities in networks.
\newblock {\em PNAS}, 101(9):2658--2663, 2004.

\bibitem{Schaeffer07_survey}
S.~Schaeffer.
\newblock Graph clustering.
\newblock {\em Comp. Sci. Review}, 2007.

\bibitem{simmel64affiliations}
G.~Simmel.
\newblock {\em Conflict and the web of group affiliations}.
\newblock Simon and Schuster, 1964.

\bibitem{tsourakis08triangles}
C.~E. Tsourakakis.
\newblock {Fast Counting of Triangles in Large Real Networks without Counting:
  Algorithms and Laws}.
\newblock In {\em ICDM '08}, 2008.

\bibitem{luxburg05_survey}
U.~von Luxburg.
\newblock A tutorial on spectral clustering.
\newblock {\em Statistics and Computing}, 17:395--416, 2007.

\bibitem{watts98collective}
D.~Watts and S.~Strogatz.
\newblock Collective dynamics of small-world networks.
\newblock {\em Nature}, 393:440--442, 1998.

\bibitem{jaewon11community}
J.~Yang and J.~Leskovec.
\newblock Structure and overlaps of communities in networks
\newblock Technical Report, Stanford Infolab, 2012.

\bibitem{zheleva09affiliation}
E.~Zheleva, H.~Sharara, and L.~Getoor.
\newblock Co-evolution of social and affiliation networks.
\newblock In {\em KDD '09}, 2009.

\end{thebibliography}


\end{document}